\documentclass[sigconf]{acmart}
\AtBeginDocument{%
  }

\usepackage{graphicx}
\usepackage{subcaption}
\usepackage{booktabs} 
\usepackage{tabularx}
\usepackage{multicol}
\usepackage{multirow}
\usepackage{hyperref}
\usepackage{tcolorbox}
\usepackage{xcolor}
\tcbuselibrary{breakable, skins}
\usepackage{listings}
\lstdefinelanguage{json}{
    basicstyle=\ttfamily\small,
    numbers=none,
    showstringspaces=false,
    breaklines=true,
    frame=single,
    string=[s]{"}{"},
    comment=[l]{//},
    morecomment=[s]{/*}{*/},
    morekeywords={true,false,null},
    escapeinside={(*@}{@*)}
}
\usepackage{placeins}
\usepackage{enumitem}
\setlist{noitemsep, topsep=3pt}

\usepackage[varqu]{zi4} 

\usepackage{tikz}
\usetikzlibrary{arrows.meta,positioning,calc}

\begin{document}

\title{Beyond Reproducibility: Towards Security-Aware Evaluation of Research Artifacts
}


\author{Nanda Rani}
\affiliation{%
  \institution{CISPA – Helmholtz Center for Information Security}
  \country{Germany}
  }
\email{nanda.rani@cispa.de}

\author{Christian Rossow}
\affiliation{%
  \institution{CISPA – Helmholtz Center for Information Security}
  \country{Germany}
  }
\email{rossow@cispa.de}








\begin{abstract}
Research artifacts are widely shared to support reproducibility, and artifact evaluation (AE) has become common at many leading conferences. However, AE mainly checks whether artifacts work as claimed and can be reproduced. It does not aim at spotting or mitigitating potential security risks. Since these artifacts are publicly released and reused, they may unintentionally create opportunities for misuse and raise concerns about safe and responsible sharing. 
We study $1{,}388$ research artifacts published between 2023 and 2025 at the top-4 security conferences, perform static analysis, and obtain $132{,}431$ candidate security findings. We propose a taxonomy for context-aware security assessment and examine the findings to filter false positives and identify findings that represent plausible context-dependent security risks.
We find that $44.80\%$ of the reviewed findings are security-relevant. 
To support scalable analysis, we present SAFE (Security-Aware Framework for Artifact Evaluation), an autonomous framework that assesses tool-reported findings based on code semantics, execution context, and practical exploitability. SAFE achieves $94.40\%$ accuracy and a $93.60\%$ F1-score in distinguishing security-relevant from non-security findings, and $92.40\%$ accuracy and an $81.10\%$ F1-score in classifying security-risk types. Overall, our results show that context-aware security assessment is a practical complement to existing AE processes and can support safer and more responsible research artifact sharing. The source code for SAFE is available  at: \url{https://github.com/nanda-rani/SAFE}
\end{abstract}



\keywords{
Research Artifacts, Artifact Evaluation, Software Security, Large Language Models, Agentic AI, Security Risk Assessment
}


\maketitle

\section{Introduction}
\label{sec:introduction}


The sharing of research artifacts has become a central part of modern scientific research practice, especially in systems and security research~\cite{olszewski2023get,wu2026agent}. Authors now routinely release code, datasets, and scripts alongside their papers to support reproducibility and transparency~\cite{malik2020artifact,olszewski2025reproducibility,sedghpour2024artifact,vansteenhuyse2026not}. Many top-tier venues encourage this practice through artifact evaluation (AE) processes and badge systems, which reward artifacts that are available, functional, and reproducible~\cite{guevara2024research,guilloteau2024artifact,olszewski2025sok,winter2022retrospective}. Some venues, such as USENIX Security, even mandate artifact release as part of their open-science requirements~\cite{usenixsecurity26}.

While this shift has improved the credibility and transparency of experimental results, current evaluation pipelines still focus mainly on whether an artifact works as expected~\cite{baek2026artisan,beyer2025artifact,olszewski2023get,winter2022retrospective}. They check if the code runs and reproduces reported results~\cite{guilloteau2024artifact,olszewski2025reproducibility}. 
Security, however, is not an explicit evaluation criterion. Existing AE processes emphasize properties such as availability, functionality, and reproducibility, without requiring a systematic assessment of security risks. Consequently, the reputation of established security venues and their review processes may implicitly carry over to perceptions of the artifacts they release, even though artifact evaluation provides no security guarantee. This leaves the community with limited evidence about the security risks present in released research artifacts.


This gap introduces significant risks in real-world reuse scenarios. Research artifacts are widely shared and reused by researchers, practitioners, and developers~\cite{heumuller2020publish}. They often include complex code, dependencies, and experimental scripts, where vulnerable dependencies can propagate security risk through software dependency networks~\cite{robinson2026original}. Even when not intended for production, parts of these artifacts are frequently reused or adapted in real systems~\cite{muttakin2026state,winter2022retrospective,zilberman2020thoughts}. Prior work has also shown that insecure code obtained from external development resources can propagate into developers' software~\cite{oh2024poisoned}.

\begin{figure}[htbp]
    \centering
    \includegraphics[width=\columnwidth]
        {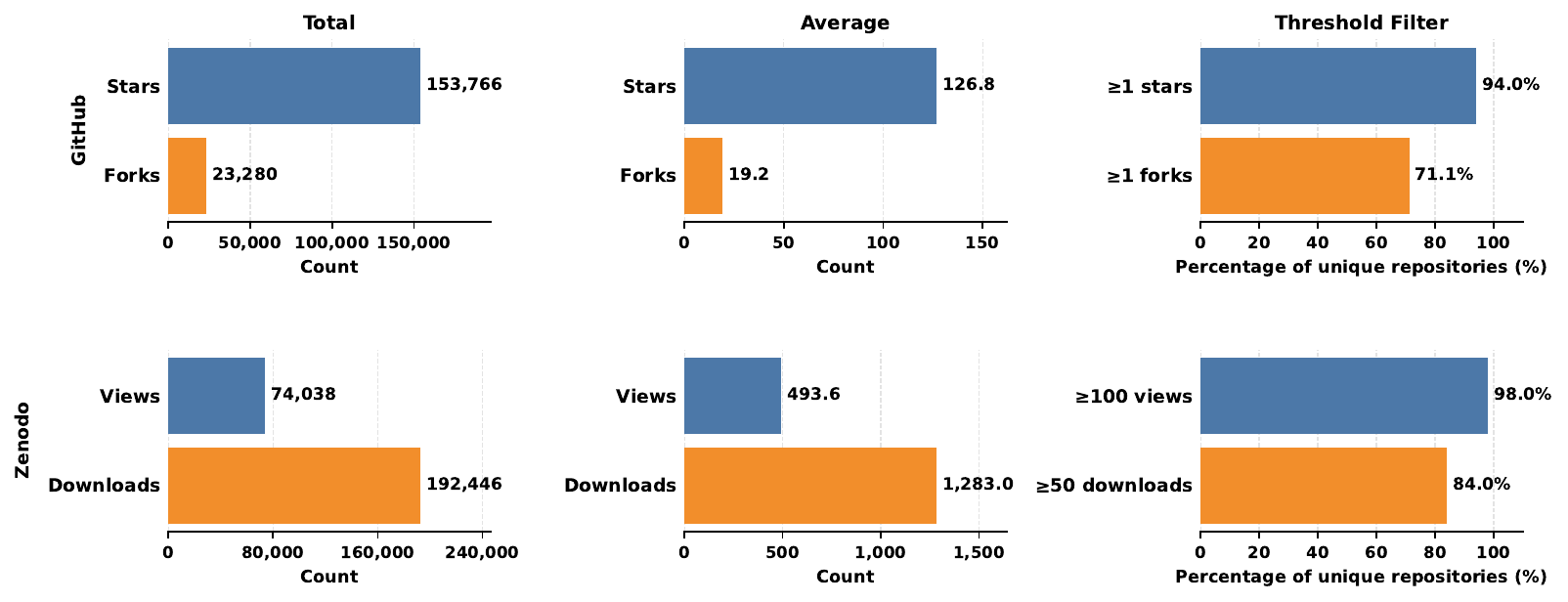}
        
    \caption{Community engagement associated with the artifacts analyzed in
    this paper.\protect\footnotemark}
    \label{fig:artifact_engagement}
    \vspace{-3mm}
\end{figure}

The potential impact of insecure research artifacts is amplified by their
substantial community engagement. As shown in
Figure~\ref{fig:artifact_engagement}, the GitHub repositories in
our curated dataset collectively received 153,766 stars and 23,280 forks, averaging
126.8 stars and 19.2 forks per repository. 
\footnotetext{Zenodo engagement may include accesses by artifact evaluation committee. We therefore report the percentages of records with at least
100 views and at least 50 downloads, rather than records with any activity.}
Engagement was also widespread:
94.0\% of these repositories received at least one star, while 71.1\% were
forked at least once. Similarly, the Zenodo records accumulated 74,038
views and 192,446 downloads, averaging 493.6 views and 1,283.0 downloads per
record. Moreover, 98.0\% received at least 100 views, and 84.0\% received at
least 50 downloads. These measurements show that research artifacts receive substantial attention and are frequently accessed, copied, or redistributed, which increases their exposure to potential security risks.

Our analysis provides a concrete example of how security risks can emerge when research artifacts are reused beyond their intended setting. We identified a publicly released artifact for software version fingerprinting that constructs shell commands using a user-supplied target host or address through \texttt{subprocess.Popen(..., shell=True)}. In its original experimental workflow, this input is assumed to be trusted. However, if the same implementation is reused in a deployed scanning service or exposed through an API without additional input validation, an attacker could inject arbitrary shell commands. This shows how implementation choices that are reasonable within a controlled research environment can become security risks under different reuse conditions. These observations motivate assessing artifact security alongside reproducibility.

The research artifacts may include insecure coding patterns such as unsafe deserialization, hardcoded secrets, vulnerable dependencies, or misconfigurations. Identifying these issues is not straightforward.  Static analysis tools can detect potential problems, but they often generate a large number of warnings without considering context~\cite{guo2023mitigating,kang2022detecting,ghosh2025wasn}. 
In addition, the presence of a flagged issue does not necessarily mean it is exploitable; its risk depends on real-world usage. 
Some findings remain harmless in controlled settings but may become exploitable upon reuse. These gaps motivate the following research questions:

\begin{enumerate}[label=\textbf{RQ\arabic*:}, leftmargin=*]
    \item To what extent do publicly released research artifacts contain security issues, and how often do these issues represent real risks in practice?

    \item How can security findings in research artifacts be analyzed in a scalable and context-aware way using an automated framework?
\end{enumerate}

To answer RQ1, we conduct an empirical study of security risks in publicly shared research artifacts from top-tier security conferences. We analyze $1{,}388$ artifacts using static analysis tools and observe that insecure practices are not uncommon. At the same time, not all tool-reported warnings lead to actual exploitability in artifact-specific contexts.
To support a more detailed analysis, we consider factors such as attacker-controlled input, execution context, and the conditions required for exploitation. Based on this, we assign context-aware security labels to the static analysis findings.
To answer RQ2, we propose SAFE (Security-Aware Framework for Artifact Evaluation), a first step toward a scalable and context-aware approach for assessing the security risk of research artifacts. The framework reasons about code semantics, execution context, and attacker feasibility. It uses a large language model (LLM) to connect raw static analysis results with practical security understanding.
SAFE also brings structured reasoning and security checklists into the evaluation process, so that security can be considered alongside reproducibility and functionality. In summary, this paper makes the following contributions:


\begin{itemize}
    \item We present an empirical study of security risks in public research artifacts by analyzing $1{,}388$ artifacts published between 2023 and 2025 at the top-4 security conferences.
     \item We introduce a taxonomy for context-aware security assessment of static findings to support structured classification.
    \item We show that insecure practices are common in public research artifacts and that tool-reported warnings do not always reflect practical security risks.
    \item We present SAFE, a framework that reasons about code semantics, execution context, and attacker feasibility for artifact-level security assessment. We evaluate it on static analysis findings and show that it distinguishes non-security findings from plausible context-dependent security risks.
\end{itemize}
\section{Background \& Related Work}
\label{sec:back_relwork}

Artifact evaluation was first introduced as a voluntary process at ESEC/FSE 2011~\cite{beyer2025artifact} and has become a common practice at most top-tier venues.
AE aims to improve transparency and reliability by encouraging authors to share artifacts such as source code, datasets, and experimental pipelines along with their papers.

Over time, AE has developed into a structured process supported by dedicated committees and incentive mechanisms such as badge systems~\cite{beyer2025artifact}. These badges recognize aspects such as artifact availability, functionality, and reproducibility~\cite{guevara2024research,hermann2022has}. As a result, sharing artifacts has become more common, and reproducibility is now seen as an essential part of scientific progress~\cite{beyer2025reproducibility,barr2023continuously}.

\subsection{Related Work}

Existing studies examine AE’s role in reproducibility and identify both its strengths and limitations.
Olszewski et al.~\cite{olszewski2023get} study artifacts from top-tier security conferences and show that AE improves reproducibility success rates. Winter et al.~\cite{winter2022retrospective} review artifact evaluation practices over the past decade and find that evaluated artifacts do not always meet expectations in terms of usability and long-term impact. Guilloteau et al.~\cite{guilloteau2024longevity} highlight concerns around artifact longevity and sustainability, which suggests that current evaluation criteria capture only a limited view of reproducibility. Vansteenhuyse et al.~\cite{vansteenhuyse2026not} conduct a 25-year longitudinal analysis of artifact sharing in top-tier cybersecurity venues and examine the impact of conference policies on artifact availability and hosting practices.

Recent work explores LLM-based agents for automated artifact evaluation and result reproduction to reduce reviewer effort.
Baek et al.~\cite{baek2026artisan} present Artisan, an LLM-based agent that generates and validates reproduction scripts to assess reproducibility at scale. Heye et al.~\cite{heye2025supporting} propose a toolkit for reproducibility assessment and environment setup to reduce reviewer effort and improve scalability. Wu et al.~\cite{wu2026agent} introduce ArtifactCopilot, an agent-based framework to automates environment setup, execution, and error recovery.
These approaches mainly focus on improving the scalability and efficiency of reproducibility checks.

Beyond artifact evaluation, empirical studies have examined security issues in software ecosystems. Ghosh et al.~\cite{ghosh2025wasn} show that explicit security labels capture only a fraction of security-related reports in open-source npm repositories, highlighting the importance of contextual analysis. Ye et al.~\cite{ye2024detecting} combine LLM reasoning with static taint analysis to detect command-injection vulnerabilities missed by conventional analysis. Robinson et al.~\cite{robinson2026original} show how vulnerabilities in a small number of npm packages can propagate to many dependent packages.



Prior work on artifact evaluation focuses mainly on reproducibility and usability. However, the security implications of publicly released research artifacts remain largely unexplored. Current AE guidelines and badge systems do not assess insecure coding practices, vulnerable dependencies, or exploitable behavior. This gap is important in security research, where artifacts may be reused in real systems and potentially carry forward hidden risks.


Our work addresses this gap through an empirical study of security risks in research artifacts. We introduce a context-aware approach that connects static analysis findings with practical exploitability. By combining empirical analysis, a structured taxonomy, and automated reasoning, we extend artifact evaluation beyond reproducibility to security.

\section{Threat Model}
\label{sec:threat_model}

We study security risks in research artifacts under realistic usage. Artifacts are often shared with good intent, but they may still include weaknesses. Our focus is on what happens after release. Artifacts may be reused, extended, or executed in shared or cloud environments. Risks can arise when code is unsafe, configurations are weak, or inputs are wrongly assumed to be trusted. Such attacker influence can arise in realistic usage. For example, users may run artifacts on datasets downloaded from public repositories, reuse shared configuration files, or integrate code into workflows where inputs are not strictly validated. In these cases, adversarial content can propagate into execution without requiring privileged access.

We assume an attacker interacts with the artifact as a normal user or through its exposed interfaces. The attacker can supply inputs such as data files, configuration values, or command-line arguments. We do not assume privileged access or direct code modification. The main risk is to users who run or reuse the artifact, where unsafe input or coding practices can affect their system or data. A vulnerability is treated as exploitable only when three conditions hold: (i) The affected code is actually executed. (ii) The attacker can control the input or path that reaches it. (iii) The attack does not rely on unrealistic setup or assumptions, such as modifying internal code or accessing non-exposed interfaces.

\section{Dataset \& Measurement Setup}
\label{sec:dataset}

This section describes the artifact collection, static analysis pipeline, and an overview of the findings. It establishes the dataset scale and diversity for subsequent contextual analysis.
\subsection{Artifact Collection}


We collected artifact metadata for 2,268 papers published at ACM CCS, NDSS, the IEEE Symposium on Security and Privacy (S\&P), and USENIX Security during 2023–2025, covering twelve conference-year collections. We obtained these records from Vahldiek-Oberwagner et al.~\cite{reprodb2026}. Of the 2,268 paper records, 352 did not pass the artifact-URL eligibility check due to the unavailability of the required field. The remaining 1,916 records entered the acquisition stage.
Of the 1,916 acquisition attempts, 1,338 Git repositories were successfully cloned, 554 artifacts were downloaded as files or archival records, and 24 attempts failed. 
The failures comprised nine DNS-resolution errors, seven HTTP 4xx client errors, three GitHub repository-not-found errors, two TLS failures, one GitLab authentication failure, one HTTP 503 response, and one read timeout, as summarized in Table~\ref{tab:download_failure_reasons} in the Appendix. 
We therefore successfully collected 1,892 artifacts.

The collected resources included source-code repositories, scripts, notebooks, datasets, model weights, documents, websites, and experimental packages. Since our analysis focuses on released implementation code, we retained artifacts with recognized source code or source-bearing archives. Of the 1,892 successfully collected artifacts, 504 contained only non-source material and were excluded. This left 1,388 code-bearing artifacts for analysis. Complete exclusion criteria and counts are provided in Table~\ref{tab:complete_exclusion_reasons} in the Appendix.

\begin{figure}[htbp]
    \centering
    \includegraphics[
        width=0.8\columnwidth
    ]{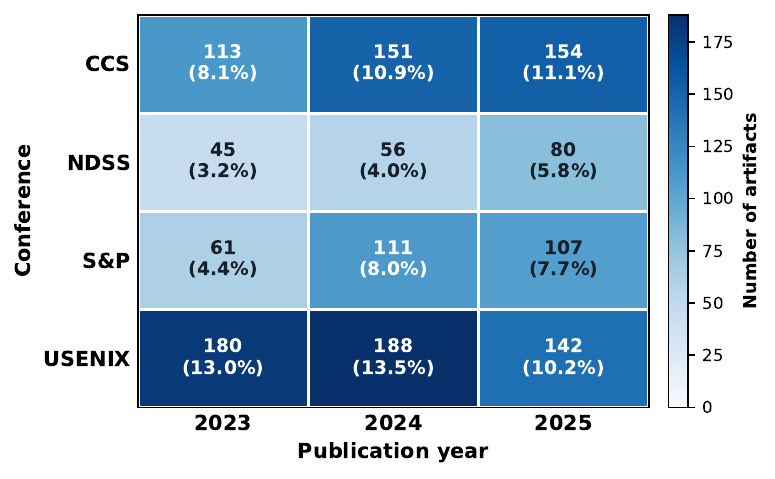}
    \caption{
        Distribution of the finalized code-bearing artifacts across conferences and publication years.
    }
    \label{fig:conference_year_distribution}
\end{figure}

The final cohort contains 1,388 code-bearing artifacts from four major security conferences between 2023 and 2025. Fig.~\ref{fig:conference_year_distribution} shows their distribution across conferences and publication years. USENIX Security contributes the largest share with 510 artifacts (36.7\%), followed by CCS with 418 (30.1\%), S\&P with 279 (20.1\%), and NDSS with 181 (13.0\%). The yearly distribution is relatively balanced: 399 artifacts are associated with 2023, 506 with 2024, and 483 with 2025.

\begin{figure}[htbp]
    \centering
    \includegraphics[width=0.85\linewidth]{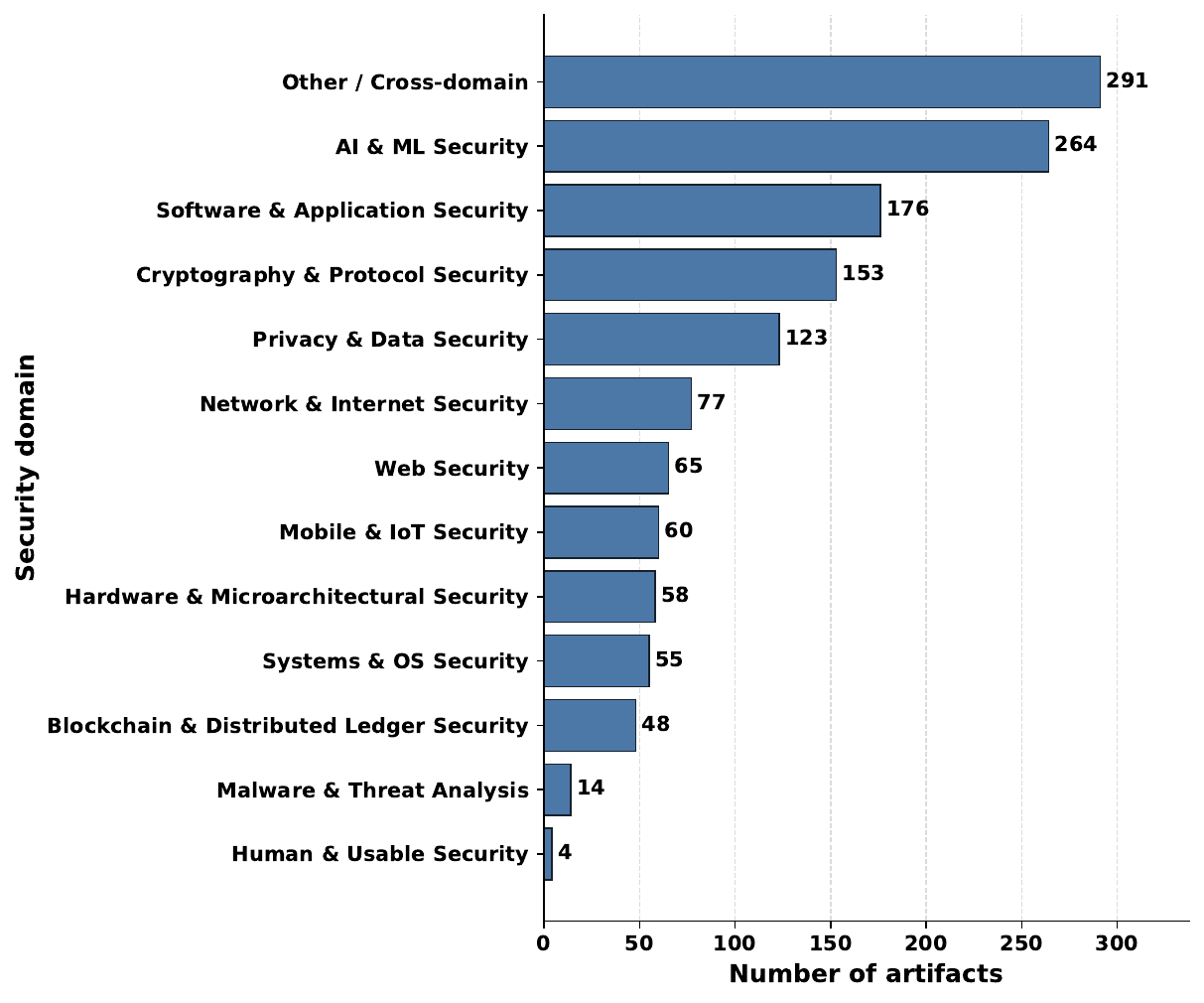}
    \caption{Distribution across security domains}
    \label{fig:security-domain}
\end{figure}

\begin{figure}
    \centering
    \includegraphics[width=0.85\linewidth]{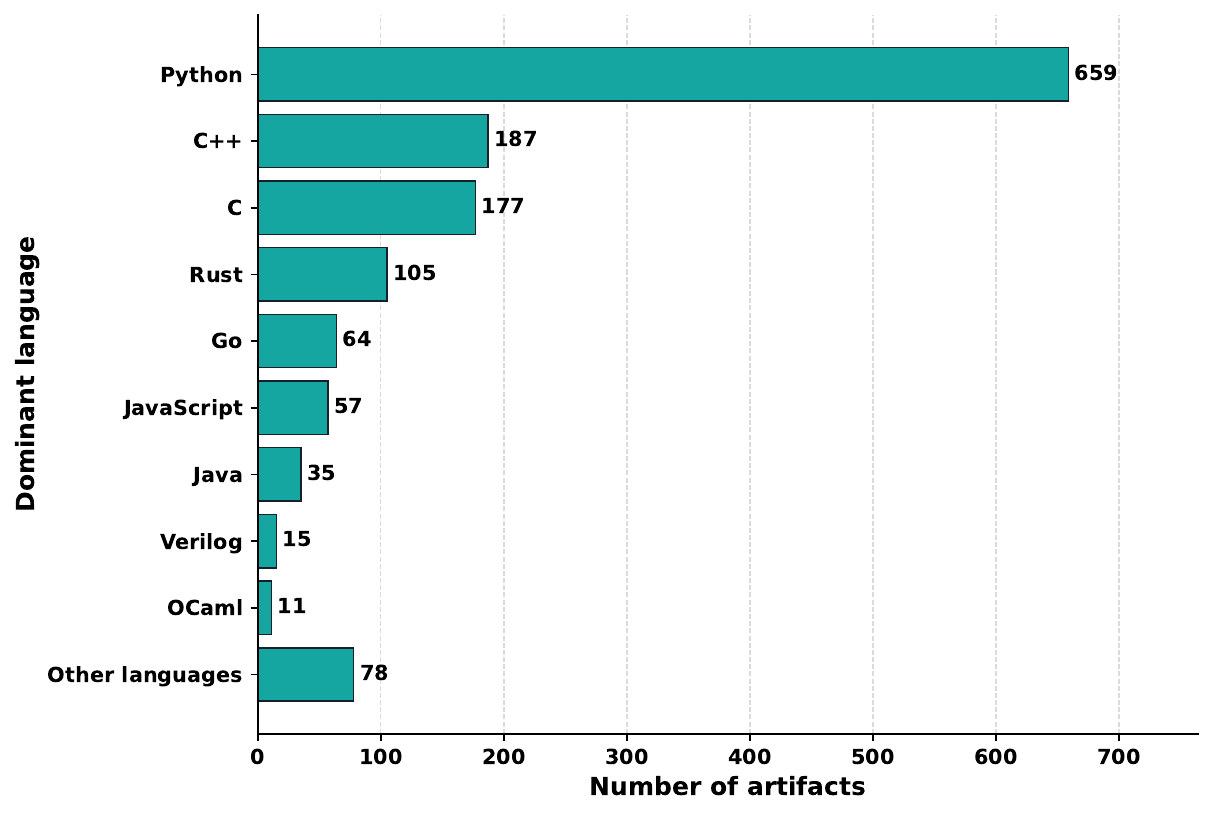}
    \caption{Distribution of the top 10 programming languages}
    \label{fig:language}
\end{figure}

The collection covers several security domains, including systems security, AI security, and software security. It also includes a wide range of programming languages, dependency setups, and execution workflows. Some artifacts consist of lightweight scripts, while others contain complex, multi-component systems. Fig.~\ref{fig:security-domain} and~\ref{fig:language} show the distribution of the 1,388 artifacts by security domain and dominant programming language, which reflects the technical diversity of our dataset.



\subsection{Static Analysis Pipeline}
\label{sec:static_analysis_pipeline}

We use a multi-tool static analysis pipeline to identify candidate security
issues in the collected artifacts. At the code level, we use
Semgrep~\cite{semgrep} to detect potentially insecure implementation patterns,
such as command injection, unsafe deserialization, cryptographic misuse, and
other rule-specific coding weaknesses. We additionally use
Trivy~\cite{trivy} in two modes: vulnerability scanning to identify
third-party dependencies associated with known security advisories, and
secret scanning to identify potential credentials, tokens, and private keys
embedded in artifact files. These analyses provide complementary coverage of
source-code weaknesses and vulnerable dependencies.

Each scanner is executed only when it is applicable to the artifact's
contents. Semgrep is applied to artifacts containing supported source-code
files, while Trivy vulnerability scanning is applied when recognizable
dependency metadata is available. Trivy secret scanning examines relevant
artifact files independently of dependency detection. For each artifact, we run both tools across the codebase and associated dependency files.  
We normalize the tool-specific outputs into a common finding representation.
Each record identifies the artifact, originating tool, finding type, reported
severity, affected file or dependency, source location, and available
vulnerability metadata. We collapse only exact duplicate scanner records,
while retaining their original multiplicity for reproducibility. 
The complete representation is described in
Appendix~\ref{app:finding_schema}.

\subsection{Initial Findings Overview}
\label{sec:initial_findings}

Across the 1,388 shortlisted artifacts, the static analysis pipeline
produced 132,929 raw scanner records. After collapsing 498 exact
duplicates while retaining their original multiplicities for auditing,
we obtained 132,431 unique candidate findings. We use the term
\emph{candidate finding} because a tool report does not, by itself,
establish that the reported weakness is reachable or exploitable in the
artifact's intended or downstream execution context.

Semgrep contributed 75,851 code-level candidates and Trivy reported
56,580 dependency-vulnerability candidates. 
Semgrep produced findings for 1,001 artifacts and trivy related
vulnerabilities were reported for 605 artifacts. These categories overlap because an artifact may contain findings from multiple scanners.

The findings were associated with 328 distinct CWE identifiers.
Because a finding can map to more than one CWE, we rank CWEs according
to the number of distinct artifacts in which they occur rather than
their raw number of occurrences. Table~\ref{tab:top_cwe} presents the
ten most prevalent CWEs. Deserialization of untrusted data
(CWE-502) was the most widespread, occurring in 611 artifacts, followed
by OS command injection (CWE-78) in 400 artifacts and the use of
potentially dangerous functions (CWE-676) in 345 artifacts.
\begin{table}[htbp]
\centering
\caption{Top CWE categories identified in candidate findings.}
\label{tab:top_cwe}
\begin{tabularx}{\linewidth}{p{1.3cm}Xr} 
\toprule
\textbf{CWE ID} & \textbf{Name} & \textbf{\# Artifacts} \\
\midrule
CWE-502 & Deserialization of Untrusted Data & 611 \\
CWE-78 & Improper Neutralization of Special Elements used in an OS Command & 400 \\
CWE-676 & Use of Potentially Dangerous Function & 345 \\
CWE-400 & Uncontrolled Resource Consumption & 322 \\
CWE-770 & Allocation of Resources Without Limits or Throttling & 301 \\
CWE-95 & Improper Neutralization of Directives in Dynamically Evaluated Code ('Eval Injection') & 293 \\
CWE-1333 & Inefficient Regular Expression Complexity & 278 \\
CWE-20 & Improper Input Validation & 276 \\
CWE-22 & Improper Limitation of a Pathname to a Restricted Directory ('Path Traversal') & 253 \\
CWE-94 & Improper Control of Generation of Code ('Code Injection') & 242 \\
\bottomrule
\end{tabularx}
\end{table}

For dependency vulnerabilities, Trivy classified 4.53\% candidates as
critical, 43.81\% as high, 38.72\% as medium, 11.50\% as low, and 1.44\% as unknown, as shown in Table~\ref{tab:severity_distribution}.
The severity of these issues ranges from low-severity warnings to critical vulnerabilities. We also observe differences  across the tools used, as Semgrep identifies pattern-based code issues, while Trivy detects vulnerabilities in third-party dependencies. They reveal that potential security issues are common across research artifacts.
\begin{table}[htbp]
\centering
\caption{Distribution of vulnerability findings by severity.}
\label{tab:severity_distribution}
\begin{tabularx}{\linewidth}{Xrrrrr}
\toprule
\textbf{Severity} & \textbf{High} & \textbf{Medium} & \textbf{Low} & \textbf{Critical} & \textbf{Unknown$^{*}$} \\
\midrule
\textbf{Ratio} & 43.81\% & 38.72\% & 11.50\% & 4.53\% &  1.44\% \\
\bottomrule
\end{tabularx}
\raggedright
\footnotesize{$^{*}$ The \emph{Unknown} category indicates that Trivy identified a vulnerability record but did not provide a severity classification from its available data sources.
}
\vspace{-3mm}
\end{table}


These initial results establish the scale and breadth of tool-reported security concerns. However, as we show in the subsequent analysis, a large fraction of these findings lack sufficient context to determine their practical exploitability. We therefore further examine which candidates remain meaningful after considering execution context and realistic reuse scenarios. This dataset enables us to systematically characterize the gap between tool-reported findings and realistic security risks across diverse research artifacts.

\vspace{-0.5mm}
\section{Taxonomy for Contextual Security Analysis} 
\label{sec:taxonomy}

To distinguish between theoretical vulnerabilities and practical risks, we introduce a taxonomy for contextual security assessment of research artifacts. It captures the relationship between static analysis findings and realistic exploitability.
\subsection{Design Principles}
\label{subsec:design_principle}
Our taxonomy follows four principles:
(i) \textbf{Exploitability over syntax:} We distinguish insecure patterns from vulnerabilities that are exploitable in practice.
(ii) \textbf{Context-aware reasoning:} Findings are interpreted based on how the artifact is used, such as training, evaluation, or demo code.
(iii) \textbf{Evidence-driven analysis:} Decisions rely on observed code behavior and dependencies, without assuming undocumented inputs or deployments.
(iv) \textbf{Conservative classification:} When evidence is limited, we avoid overestimating risk and account for uncertainty.


\subsection{Contextual Dimensions}
\label{subsec:contextual_dimension}
To enable consistent and meaningful classification, we define a set of dimensions that capture how each finding behaves in its actual usage context. These dimensions help us reason about whether a reported issue can translate into a practical risk. We consider the following dimensions while analyzing the findings:

\begin{itemize}

\item \textbf{Attacker-controlled input:} This dimension captures whether an adversary can influence inputs reaching the flagged code in the artifact's current or plausible reuse context. Examples include network request parameters, uploaded files, externally supplied URLs or hostnames, and data obtained from untrusted sources. It is labeled \texttt{yes}, \texttt{no}, or \texttt{uncertain}, with a brief explanation.

\item \textbf{Reachability:} This dimension checks whether the flagged code is executed during use. If it is unreachable, the impact is limited. The label is \texttt{yes}, \texttt{no}, or \texttt{uncertain}, with a brief note on whether an attacker can trigger it.

\item \textbf{Execution context:} This dimension describes the role of the code within the artifact. For example, a training pipeline, evaluation script, demo, or deployment component. This context affects exposure and risk. The label is expressed as text, explaining where and how the code runs in the workflow.

\item \textbf{Exploitation condition:} This dimension outlines the conditions needed to exploit the issue, such as vulnerable dependencies, malicious inputs, or specific privileges. The description helps clarify whether exploitation is straightforward or requires strong assumptions.

\end{itemize}

\noindent These dimensions serve as necessary conditions for practical exploitability. Without attacker-controlled input and reachable execution, a reported issue is unlikely to manifest as a real security risk. Together, they provide a structured basis for interpreting tool-reported findings. This enables us to move beyond surface-level warnings and assess the actual risk posed by research artifacts.

\subsection{Classification Labels}
\label{subsec:classification_label}
As false positives are a well-known issue in static analysis of codebases, we first assign a category to each finding before determining its security label. The findings are divided into two categories: (i) \textit{security-relevant findings} and (ii) \textit{non-security findings}.
If a finding reported by static analysis tools is identified as a false positive, it is directly categorized as a \textit{non-security finding}. In contrast, true positives are categorized as \textit{security-relevant findings}. The formal taxonomy categorization flow is illustrated in Figure~\ref{fig:taxonomy_flow}. 
\begin{figure}[htbp]
    \centering
    \vspace{-1mm}
    \definecolor{boxBlue}{RGB}{218,232,252}
    \definecolor{boxYellow}{RGB}{255,229,153}
    \definecolor{boxTeal}{RGB}{154,199,191}
    \definecolor{boxRed}{RGB}{255,153,153}
    \definecolor{boxOrange}{RGB}{255,204,153}
    \definecolor{boxIndigo}{RGB}{153,153,255}
    \resizebox{\linewidth}{!}{%
    \begin{tikzpicture}[
    every node/.style={
        rectangle, rounded corners=7pt, draw=black, line width=1.4pt,
        font=\bfseries\sffamily, align=center,
        minimum height=1cm, inner sep=3pt, anchor=west
    },
    conn/.style={
        line width=1.4pt, black, -{Triangle[length=3.2mm,width=3mm]}
    }
]
 
\node[fill=boxBlue, text width=2.3cm] (root) at (0,2.15) {Static\\ Analysis\\ Findings};
 
\node[fill=boxYellow, text width=2.4cm] (sec) at (4.6,2.95) {Security-\\relevant\\ Findings};
\node[fill=boxTeal,   text width=2.4cm] (nonsec) at (4.6,1.05) {Non-security\\ Findings};
 
\node[fill=boxRed]    (ctx)  at (8.6,3.55) {CONTEXTUAL\_RISK};
\node[fill=boxOrange] (hard) at (8.6,2.25) {HARDENING\_RECOMMENDATION};
\node[fill=boxIndigo] (fp)   at (8.6,1.05) {FALSE\_POSITIVE};
 
\draw[conn] (root.east) -| ($(root.east)!0.5!(sec.west)$) |- (sec.west);
\draw[conn] (root.east) -| ($(root.east)!0.5!(nonsec.west)$) |- (nonsec.west);
 
\draw[conn] (sec.east) -| ($(sec.east)!0.5!(ctx.west)$)  |- (ctx.west);
\draw[conn] (sec.east) -| ($(sec.east)!0.5!(hard.west)$) |- (hard.west);
 
\draw[conn] (nonsec.east) -- (fp.west);
 
\end{tikzpicture}%
    }
    \caption{Overview of the formal taxonomy categorization}
    \label{fig:taxonomy_flow}
\end{figure}
We assign security labels to each relevant finding using our taxonomy, guided by the design principles in Sections~\ref{subsec:design_principle} and~\ref{subsec:contextual_dimension}. 
These labels capture the practical relevance of a reported issue, rather than merely its presence in the code. The labels used are described below:

\begin{itemize}


\item \texttt{CONTEXTUAL\_RISK:} This label captures issues that could become exploitable depending on how the artifact is used. The risk may arise under real-world deployment settings. For example, a command injection flaw in a script may become dangerous if the same code is later exposed through an API.

\item \texttt{HARDENING\_RECOMMENDATION:} This label refers to unsafe practices with low practical impact in the current context. They are not immediate threats but should be improved for better security. For example, weak cryptography in non-critical components or missing validation in internal utilities.          


\item \texttt{FALSE\_POSITIVE:} This label is used when the reported issue does not correspond to a real problem in the given context. This happens due to incorrect pattern matching or when the flagged code is not relevant to execution.

\end{itemize}

\noindent These labels structure the interpretation of tool-reported findings.
The taxonomy connects raw findings to practical risk assessment and supports our subsequent analysis.                                                          

\section{Empirical Analysis and Key Findings (RQ1)}
\label{sec:manual_analysis}

Static analysis alone does not establish whether a weakness is reachable, exploitable, or relevant to the artifact's execution context. We therefore apply our contextual security assessment taxonomy to a structured subset of the candidate findings.

The pipeline produced 132,431 unique candidate findings. These
candidates represented 4,194 distinct \emph{finding types}, where a
finding type is defined as the combination of the reporting tool and its
finding identifier. For Semgrep, the identifier corresponds to a rule
identifier, whereas for Trivy dependency scanning it corresponds to a
vulnerability identifier. Tool qualification prevents identifiers from
different scanners from being treated as the same type.

Because a single large artifact may contain many occurrences of the
same finding type, we measure prevalence using the number of distinct
artifacts containing the type rather than its raw number of occurrences.
We identified 287 finding types occurring in at least 30 distinct
artifacts. From these, we selected the 50 types with the highest
artifact-level prevalence. The least prevalent selected type occurred
in 85 artifacts.

For each selected finding type, we sampled one candidate from each of
five distinct artifacts, producing 250 instances for contextual
assessment. 
The resulting sample spans 176
distinct artifacts. 
The sample-construction counts and conference-year composition are
reported in Tables~\ref{tab:contextual_sample} and
\ref{tab:contextual_sample_distribution}, respectively, in
Appendix~\ref{app:contextual_sampling}.

This process provides balanced coverage of recurring finding types:
each selected type contributes the same number of reviewed instances.
It is therefore suitable for examining how the contextual relevance of
a recurring pattern varies across artifacts. 

\paragraph{Overall Analysis.}
Static analysis tools report candidate security findings rather than confirmed
vulnerabilities. Their outputs therefore require contextual assessment,
particularly for research artifacts whose execution assumptions may differ
from those of deployed software. For each sampled finding, we examined whether the reported construct or affected dependency was actually present, whether the relevant code was reachable, whether less-trusted input could influence it, and how it was used within the artifact workflow.

We classify a finding as \texttt{FALSE\_POSITIVE} only when the condition underlying the tool's security warning is absent in the inspected artifact. This may occur, for example, when the reported input cannot originate from an untrusted source, the flagged code path is unreachable, or the vulnerable behavior identified by the tool is not exercised in the artifact's execution context. The absence of a demonstrated exploitation path alone is not sufficient for this classification. For example, one artifact was flagged because a dynamic value was passed to \texttt{urllib}. We classified this finding as \texttt{FALSE\_POSITIVE} because all observed callers obtained the URL from a fixed, checked-in mapping of model-download URLs, leaving no path for an external source to influence the value. In contrast, when such a security condition is present but exploitation depends on a different deployment or reuse context, we retain the finding as security-relevant rather than classifying it as a false positive.

As shown in Table~\ref{tab:finding_distribution}, 138 of the 250 assessed findings
(55.20\%) were classified as \texttt{FALSE\_POSITIVE}. The remaining 112
findings (44.80\%) corresponded to valid security-relevant patterns or affected
dependencies. Importantly, this does not mean that all 112 findings represent
exploitable vulnerabilities. Our contextual labels distinguish demonstrated
or plausible reuse-dependent risks from preventive hardening opportunities and
behavior that is acceptable within a bounded research workflow.
From the perspective of responsible public release of research artifacts, all labels except \texttt{FALSE\_POSITIVE} indicate potential concern and therefore warrant explicit declaration during artifact evaluation or public release.

\begin{table}[htbp]
\centering
\caption{Overall distribution of the sampled findings.}
\label{tab:finding_distribution}
\begin{tabular}{lr}
\toprule
\textbf{Finding category} & \textbf{Count} \\
\midrule
False positive & 138 \\
Security relevant & 112 \\
\midrule
Total & 250 \\
\bottomrule
\end{tabular}
\vspace{-3mm}
\end{table}
\begin{table}[htbp]
\centering
\caption{Distribution of the security-relevant findings.}
\label{tab:security_relevant_distribution}
\begin{tabular}{lr}
\toprule
\textbf{Contextual label} & \textbf{Count} \\
\midrule
\texttt{CONTEXTUAL\_RISK} & 10 \\
\texttt{HARDENING\_RECOMMENDATION} & 102 \\
\midrule
Total & 112 \\
\bottomrule
\end{tabular}
\vspace{-3mm}
\end{table}

\subsection{Analysis of Security-relevant Findings.}
Table~\ref{tab:security_relevant_distribution} shows that 102 of the 112 security-relevant findings were classified as \texttt{HARDENING\_RECOMMENDATION}. These findings identify potentially risky constructs for which we could not establish that untrusted input can reach and trigger the reported security-sensitive behavior under the artifact's current usage. Examples include unsafe deserialization interfaces and dynamic execution constructs that are not exposed to untrusted input in the observed workflow. We therefore do not treat these findings as confirmed vulnerabilities; rather, they identify preventive improvements that could reduce risk if the artifact is reused or deployed in a different setting.

10 findings were classified as \texttt{CONTEXTUAL\_RISK}. In these cases, the flagged operation was reachable and our analysis identified a concrete or strongly supported path through which less-trusted input could reach the security-sensitive operation under a realistic reuse scenario. The identified cases include deserialization, shell-command construction, dynamic evaluation, and execution of dynamically supplied code. These operation types may also appear among \texttt{HARDENING\_RECOMMENDATION} findings; the distinction is that \texttt{CONTEXTUAL\_RISK} requires a plausible path from less-trusted input to the flagged operation. For example, one artifact passes a user-supplied target hostname or IP address into a shell command that is executed using \texttt{subprocess.Popen(..., shell=True)}. Although the artifact's experimental workflow assumes this input to be trusted, reuse in a setting where the target value can be supplied by a less-trusted user could introduce command injection.

\subsection{Representative Cases}
\label{subsec:represent_cases}
The hardest judgment call in our taxonomy is separating
\texttt{HARDENING\_RECOMMENDATION} from \texttt{CONTEXTUAL\_RISK}, because both start from the same unsafe piece of code. In this section, we discuss two pairs of findings from the collected artifact runs. In each pair, the two findings use the exact same unsafe construct, but only one of them is exposed to something outside the author's control. We explain each case in four parts: what the flaw is, what the code does, a step-by-step account of the risky version, and a step-by-step account of the version that is not risky. Table~\ref{tab:case_comparison} summarizes  both pairs along the same four questions: who can influence the flagged code, whether that influence occurs under normal use, what has to go right for it to matter, and how severe the outcome is if it does.

\paragraph{Case 1: Unsafe Deserialization.}
\textbf{What the flaw is.} Python's \texttt{pickle} format does not just store data - loading a pickle file can run arbitrary code, because the format allows it to call back into Python to reconstruct objects. Calling \texttt{pickle.load} (or PyTorch's \texttt{torch.load}, which uses pickle internally) on a file is therefore equivalent to running whatever code the file's creator put there. This is a well-known weakness and listed as CWE-502.

\textbf{What the code does.} Both findings use the same shortcut: instead of recomputing an expensive result every time, the program saves it to disk with \texttt{pickle}, and on the next run it checks whether a cache file already exists and loads it instead of redoing the work. To avoid using a stale cache after the input changes, the program also stores a checksum (an MD5 hash) of the original input alongside the cached result, and compares it on the next run.

\textbf{The risky version, step by step.}
\begin{enumerate}
  \item The tool is a verifier that takes a specification file as input
    (for example, a benchmark file describing a property to check).
  \item The first time it sees a given specification file, it parses it
    and writes the parsed result to a cache file next to it, along with
    the MD5 checksum of the specification file.
    \item On a later run, the tool opens the cache file and calls
\texttt{pickle.load} before verifying its checksum. Since verification
occurs only after loading, it cannot prevent malicious deserialization.
  \item Specification and benchmark files of this kind are routinely
    shared between research groups, for example as a public benchmark
    suite. Anyone who hands over a specification file bundled together
    with a matching cache file (with a checksum that matches that
    specification file) can make their cache file's contents run on
    whoever loads it.
  \item Result: a third party who never touches the verifier's own source
    code, and does not need to break anything, can get their code to run
    on another user's machine simply by publishing a benchmark
    file that looks ordinary.
\end{enumerate}

\textbf{The hardening-only version, step by step.}
\begin{enumerate}
  \item A second tool uses the same approach: caching results with \texttt{pickle} and reloading them using a checksum.
  \item Here, the cache filename is fixed by the program and is independent of user-supplied files.
  \item We checked every place in the code that touches this cache file.
    The only one that is ever actually called is the one that
    \emph{writes} the cache. The code path that would \emph{read} an
    externally supplied cache file is never invoked in the tool's normal
    use.
  \item Result: In the documented workflow, external input cannot reach the unsafe deserialization operation. The vulnerable code exists but is not exposed to untrusted input.
\end{enumerate}

\begin{table*}[t]
\centering
\caption{Comparison of the both case pairs along the same four questions.
}
\label{tab:case_comparison}
\begin{tabularx}{\linewidth}{@{}p{1.8cm}lp{3.1cm}p{3.5cm}p{3.3cm}p{3.1cm}@{}}
\toprule
\textbf{Cases} & \textbf{Version} & \textbf{Who can supply the input?} & \textbf{Reached in normal use?} & \textbf{What has to go right?} & \textbf{What happens if it does?} \\
\midrule
\multirow{2}{1.8cm}{Unsafe\\ deserialization}
 & Risky            & Anyone distributing a benchmark file           & Yes, on every run against a shared input        & Attacker bundles a matching cache file alongside it & Arbitrary code runs on the victim's machine \\
 & Hardening   & No one outside the program itself              & No -- the read path is never called              & --- & --- \\
\cmidrule(lr){1-6}
\multirow{2}{1.8cm}{CI\\ supply chain}
 & Risky            & The upstream action's maintainer or attacker
 & Yes -- fires on every ordinary run               & Upstream action gets compromised
 & Attacker code runs with the CI job's privileges \\
 & Hardening   & Same, in principle                             & No -- guarded to the upstream repo, or never called & --- & --- \\
\bottomrule
\end{tabularx}
\end{table*}

\paragraph{Case 2: Trusting a CI Dependency at Build Time.}
\textbf{What the flaw is.} GitHub Actions workflows can reference a
third-party action either by a fixed commit hash (which never changes) or
by a mutable tag such as \texttt{@main} or \texttt{@stable} (which
always points to whatever that repository's owner most recently pushed).
When a workflow uses a mutable tag, it is trusting that repository's
owner not just today, but on every future run -- because the code that
tag points to can change after the workflow was written, without anyone
touching the artifact's own repository at all.
This is a supply-chain weakness where an attacker can target a dependency automatically used by the CI pipeline rather than the artifact itself.

\textbf{What the code does.} Both findings are CI workflow files
(\texttt{.github/workflows/*.yml}) that pull in a third-party action
using a mutable tag instead of a pinned commit hash.

\textbf{The risky version, step by step.}
\begin{enumerate}
  \item The workflow runs automatically, for example, on every push, or
    on a schedule, with nothing that would prevent it from firing.
  \item As part of that run, it pulls in a third-party action using a mutable tag, and by that point earlier steps in the job have already created files, or the job already has access to repository secrets or tokens.
  \item If the third-party action is compromised or its tag is redirected to malicious code, the workflow may execute that code during its next run.
  \item Result: whatever access that CI job has -- to secrets, to the
    checked-out repository, to publishing permissions -- is now available
    to code the artifact's authors never wrote and never reviewed, and
    the artifact's own repository was never touched to make this happen.
\end{enumerate}

\textbf{The hardening-only version, step by step.}
\begin{enumerate}
  \item A second workflow file references a third-party action the same
    way, with the same kind of mutable tag.
  \item However, this workflow is guarded so that it only runs on the
    canonical upstream project's own repository, not on copies or forks
    of it -- so it never actually executes as part of using this
    artifact. (In another case we looked at, the workflow file belongs to
    unrelated, deeply nested third-party build infrastructure that
    nothing in the artifact's documented build or test process ever
    calls.)
  \item Result: the same unpinned dependency exists in the repository,
    but no path in the artifact's documented use ever triggers that
    specific workflow to run.
\end{enumerate}



\noindent
Across all cases, the same scanner rule flagged an unsafe construct, but the actual risk differed across artifacts. We examine the code and documentation and understand how each construct was used. This shows that rule identity alone is not enough to determine practical security risk.
\begin{tcolorbox}[colback=gray!8, colframe=black, boxrule=0.5pt,
                   title=\textbf{RQ1 Takeaway}, fonttitle=\bfseries,
                   left=2mm, right=2mm, top=1mm, bottom=1mm]

Static analysis findings alone are not sufficient to determine the practical security risk of research artifacts. While security-relevant issues are common, whether they constitute real risk depends on how the flagged construct is used within the artifact. This makes context-aware assessment necessary to distinguish security-risky findings from those that primarily represent opportunities for security hardening.
\end{tcolorbox}

\section{Towards Security-Aware AE Framework}
\label{sec:llm_framework}





To automate the gap identification between findings and realistic exploitability, we present SAFE (Security-Aware Framework for Artifact Evaluation),
a first step toward a context-aware security assessment that reasons about execution context, input control, and exploit feasibility. The framework is designed to approximate analyst-style reasoning through structured, context-aware evaluation to enable scalable and automated analysis.
The framework is guided by principles that reflect how a security analyst would approach the problem in practice. It reasons about findings by considering the purpose of the code, how it is executed, and what an attacker can realistically control. It also takes into account the role of each component within the artifact to distinguish between exposed and non-exposed parts of the system. The framework also remains conservative in its decisions, explicitly acknowledges uncertainty and avoids overestimation of risk when evidence is limited, as discussed in Section~\ref{subsec:design_principle}.

\subsection{System Overview}
\begin{figure}[htbp]
    \centering
    \includegraphics[width=\linewidth]{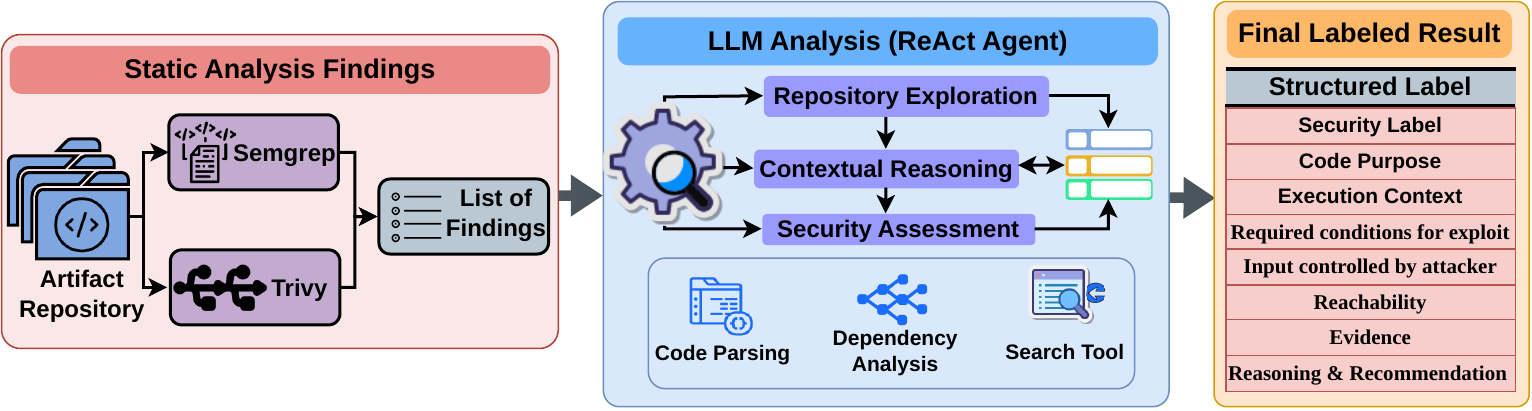}
    \caption{SAFE Workflow}
    \label{fig:safe}
    \vspace{-3mm}
\end{figure}

The framework overview is illustrated in Figure~\ref{fig:safe}.
It operates on the artifact codebase and static analysis findings generated by analysis tools.
For each finding, the framework extracts relevant contextual information, including the associated code snippet, file location, and metadata such as vulnerability type, severity, dependency, and flagged file details.
The system analyzes the surrounding code to determine its purpose, role within the artifact, dependencies, and interactions with other components.

Building on this context, the framework applies structured reasoning to assess exploitability. In our implementation, this reasoning is performed using a LLM, which interprets code semantics and integrates multiple contextual signals to produce a classification aligned with the taxonomy defined in Section~\ref{sec:taxonomy}. The LLM is instantiated in a ReAct-style agent setting~\cite{yao2022react}, where it iteratively performs reasoning and tool-assisted actions to dynamically gather additional context and evidence for making a decision. The framework augments the agent with a set of 20 read-only tools shown in the Appendix Table~\ref{tab:tools}.  


Tool augmentation allows the framework to analyze artifacts in context, understand their execution environment and dependencies, and assess exploitability in a structured manner. The toolset supports analysis at multiple levels. Repository-level tools capture the overall structure, file-level tools provide access to relevant code and configuration details, and semantic tools reason about execution flow and program behavior. These components allow the system to perform more grounded and context-aware security assessment.

The framework supports two independent classification tasks. One is a direct binary classification: \texttt{SECURITY\_RELEVANT} or \texttt{NON\_SECURITY}. The other is the multi-class taxonomy introduced in Section~\ref{sec:taxonomy}. A run can also use one of two analysis modes: zero-shot or agentic. In zero-shot mode, the vulnerability and relevant artifact files/code are provided to the LLM instance in a single pass. In agentic mode, the LLM agent can explore the repository and interact with the codebase over multiple rounds. The classification task and analysis mode are independent, allowing either analysis mode to be paired with either classification task.

\begin{table*}[htbp]
    \centering
    \caption{Performance of SAFE across binary and multi-class classification tasks under zero-shot and agentic analysis modes.}
    \label{tab:safe_performance}
    \begin{tabular}{lllcccc}
        \toprule
        \textbf{Task} & \textbf{Mode} & \textbf{Model} &
        \textbf{Accuracy (\%)} & \textbf{Precision (\%)} & \textbf{Recall (\%)} & \textbf{F1-score (\%)} \\
        \midrule
        \multirow{6}{*}{Binary class}
        & \multirow{3}{*}{Zero-shot}
        & Claude Sonnet 5 & 81.5 & 76.6 & 84.8 & 80.5 \\
        & & GPT-5.6       & 76.8 & 67.1 & 94.6 & 78.5 \\
        & & GLM-5.2       & 66.0 & 58.8 & 80.4 & 67.9 \\
        \cmidrule(lr){2-7}
        & \multirow{3}{*}{Agentic}
        & Claude Sonnet 5 & \textbf{94.4} & 95.3 & 91.9 & \textbf{93.6} \\
        & & GPT-5.6       & 88.8 & 96.7 & 77.7 & 86.1 \\
        & & GLM-5.2       & 59.6 & 53.3 & 79.5 & 63.8 \\
        \midrule
        \multirow{6}{*}{Multi-class}
        & \multirow{3}{*}{Zero-shot}
        & Claude Sonnet 5 & 78.6 & 85.7 & 64.0 & 68.3 \\
        & & GPT-5.6       & 74.0 & 72.9 & 68.2 & 68.3 \\
        & & GLM-5.2       & 61.6 & 65.3 & 49.9 & 52.1 \\
        \cmidrule(lr){2-7}
        & \multirow{3}{*}{Agentic}
        & Claude Sonnet 5 & \textbf{92.4} & 94.7 & 76.2 & \textbf{81.1} \\
        & & GPT-5.6       & 86.0 & 79.7 & 85.8 & 80.9 \\
        & & GLM-5.2       & 56.0 & 62.3 & 46.6 & 48.0 \\
        \bottomrule
    \end{tabular}
\end{table*}

\subsection{Decision Logic}

The core of the framework lies in how it reasons about each finding in a structured manner. It evaluates findings by considering  aspects that influence practical exploitability.  
It reasons about whether an external adversary can control input reaching the vulnerable code. 
When the input to the vulnerable code is not controlled by an adversary, the likelihood of exploitation is significantly reduced.

The framework also examines whether the flagged code is actually executed during normal artifact usage. Code that resides in unused paths is less likely to pose a real threat, even if it appears vulnerable in isolation.  
In addition, the broader usage context of the code is taken into account. For instance, code used only in offline experimentation has a different risk profile compared to code that may be part of a deployable system. This context helps determine how exposed the code is in practice. 
\begin{figure}[b]
    \centering
    \includegraphics[width=\linewidth]{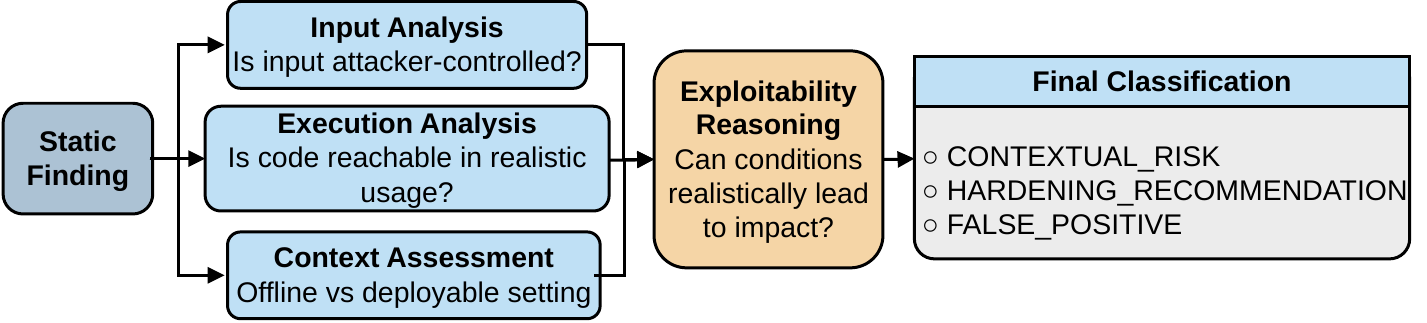}
    \caption{Decision logic workflow for assessing the exploitability of static findings. 
    }
    \label{fig:decision}
\end{figure}
The workflow shown in Figure~\ref{fig:decision} represents the structured reasoning process followed by the system. The analysis is performed iteratively by the LLM agent, which gathers evidence from the repository and evaluates multiple factors jointly before arriving at a final decision.
The agent is guided by a structured prompt specifying the classification criteria, evidence requirements, available tools, and expected output format. The complete prompts used in our experiments are included in the released research artifact (Appendix~\ref{appxsec:open_science}).

For each analyzed finding, the framework produces a structured JSON output capturing the classification, its supporting reasoning, and a remediation recommendation. The output includes the security label, code purpose, execution context, exploit conditions, evidence, reasoning, and recommendations.
Such structured outputs improve interpretability and facilitate downstream analysis. An example is shown in Listing~\ref{lst:safe_output}.

\begin{lstlisting}[language=json,caption={Example SAFE output for a finding},label={lst:safe_output}]
{
  (*@{\color{blue}"security\_label":}@*) "FALSE_POSITIVE",
  (*@{\color{blue}"code\_purpose":}@*) "....LLM training, poisoning....workflows.",
  (*@{\color{blue}"execution\_context":}@*) "requests==2.29.0 is listed in requirements.txt....the provided source files.",
  (*@{\color{blue}"required\_conditions\_for\_exploit":}@*) "A real exploit....use....insecure certificate-verification pattern ....",
  (*@{\color{blue}"input\_controlled\_by\_attacker":}@*) "no - no Requests....found in the provided code....",
  (*@{\color{blue}"reachable\_in\_artifact\_execution":}@*) "no - repository inspection found....no requests.get, requests.post....usage.",
  (*@{\color{blue}"evidence\_snippet":}@*) "requirements.txt pins requests==2.29.0....anywhere else in the codebase.",
  (*@{\color{blue}"reasoning":}@*) "The affected....requires a reachable code path....no direct....usage in the repository....",
  (*@{\color{blue}"recommendation":}@*) "No artifact-level fix....Remove unused requests....upgrade it if future code starts using..."
}
\end{lstlisting}



\noindent By integrating static analysis with context-aware reasoning, the proposed framework addresses a fundamental limitation of tool's inability to distinguish between theoretical and practically exploitable issues. 
As a first step toward security-aware artifact evaluation, SAFE's modular design supports alternative reasoning models and additional contextual signals.

\section{Evaluation}
\label{sec:eval}

We evaluate SAFE against
to measure classification performance and its ability to distinguish realistic security risks from non-exploitable or context-dependent issues.

\vspace{-1mm}
\subsection{Experiments}


To evaluate SAFE, we use the annotated subset of findings discussed in the Section~\ref{sec:manual_analysis}. This subset captures diverse categories of issues, including high-severity warnings, common insecure coding patterns, and findings occurring in different parts of the artifact. 
Each finding in this subset is labeled based on the taxonomy discussed in Section~\ref{sec:taxonomy}.
We conduct experiments in both zero-shot and agentic modes using three frontier models: \texttt{Claude-Sonnet-5}, \texttt{GPT-5.6}, and \texttt{GLM-5.2}.
For each finding, SAFE is provided with the codebase and tool-reported information, and produces both a classification and a structured explanation.
This allows us to analyze how the framework scales and to derive dataset-level insights about the nature of security issues present in research artifacts.




\begin{table*}[htbp]
    \centering
    \caption{Class-wise performance across three frontier models and two analysis modes.}
    \label{tab:safe_classwise}
    \begin{tabularx}{\linewidth}{lllccc ccc ccc}
        \toprule
        & & &
        \multicolumn{3}{c}{\textbf{Contextual\_Risk}} &
        \multicolumn{3}{c}{\textbf{Hardening\_Recommendation}} &
        \multicolumn{3}{c}{\textbf{False\_Positive}} \\
        \cmidrule(lr){4-6} \cmidrule(lr){7-9} \cmidrule(lr){10-12}
        \textbf{Mode} & \textbf{Model} & &
        \textbf{Precision} & \textbf{Recall} & \textbf{F1-score} &
        \textbf{Precision} & \textbf{Recall} & \textbf{F1-score} &
        \textbf{Precision} & \textbf{Recall} & \textbf{F1-score} \\
        \midrule
        \multirow{3}{*}{Zero-shot}
        & Claude Sonnet 5 & & 100.0 & 30.0 & 46.2 & 70.8 & 83.3 & 76.6 & 86.3 & 78.7 & 82.3 \\
        & GPT-5.6         & &  62.5 & 50.0 & 55.6 & 62.7 & 92.2 & 74.6 & 93.5 & 62.3 & 74.8 \\
        & GLM-5.2         & &  66.7 & 20.0 & 30.8 & 52.0 & 75.5 & 61.6 & 77.3 & 54.4 & 63.8 \\
        \midrule
        \multirow{3}{*}{Agentic}
        & Claude Sonnet 5 & & 100.0 & 40.0 & 57.1 & 90.3 & 92.1 & 91.2 & 93.7 & 96.4 & 95.0 \\
        & GPT-5.6         & &  60.0 & 90.0 & 72.0 & 94.7 & 69.6 & 80.2 & 84.4 & 97.8 & 90.6 \\
        & GLM-5.2         & &  66.7 & 20.0 & 30.8 & 47.9 & 76.5 & 58.9 & 72.3 & 43.5 & 54.3 \\
        \bottomrule
    \end{tabularx}
\end{table*}

\subsection{Results}
\label{subsec:results}
\subsubsection{Overall Analysis.}
\label{subsubsec:overall_analysis_result}
Table~\ref{tab:safe_performance} reports SAFE's performance across three frontier models (Claude Sonnet~5, GPT-5.6, and GLM-5.2) under two analysis modes (zero-shot and agentic) for both the binary security-relevance task and the finer-grained multi-class task. Agentic Claude Sonnet~5 is the strongest configuration overall, achieving 94.4\% accuracy and a 93.6\% F1-score on the binary task, and 92.4\% accuracy with an 81.1\% macro F1-score on the multi-class task. GPT-5.6 shows the next-largest improvement from agentic exploration, with its binary F1-score increasing from 78.5\% in zero-shot mode to 86.1\% in agentic mode. These improvements suggest that grounding classifications in evidence gathered iteratively from the repository, such as confirmed package versions, traced call sites, or verified advisories, can help models make more accurate decisions rather than relying on incomplete contextual information. Both Claude Sonnet~5 and GPT-5.6 use the additional evidence from agentic exploration to improve classification performance.

GLM-5.2 does not exhibit the same trend. Its binary F1-score decreases from 67.9\% in zero-shot mode to 63.8\% in agentic mode, while its multi-class macro F1-score decreases from 52.1\% to 48.0\%, yielding the lowest performance among the three models in both settings. We inspect and identifies that this behavior partly arises from inconsistencies between the model's reasoning and its final classification. For example, GLM-5.2 sometimes assigns \textsc{Hardening\_Recommendation} even when its generated reasoning concludes that the finding has no meaningful security relevance, such as scanner matches involving code-quality conventions or serialization calls incorrectly interpreted as deserialization. Such mismatches between the reasoning trace and the final label contribute to lower classification performance.
These results show that agentic exploration can substantially improve classification for models that effectively exploit repository-level evidence.
\subsubsection{Class-wise Analysis.}
\label{subsubsec:class_wise_label}
Table~\ref{tab:safe_classwise} breaks the multi-class results down by label. \textsc{Contextual\_Risk} is the rarest class in the ground truth and the hardest to recall across all three models: Claude Sonnet~5 attains perfect precision in both modes (100.0\%) but only recovers 30-40\% of true instances, indicating that when it does assign this label the decision is highly reliable, but it remains conservative overall. In contrast, GPT-5.6 increases recall from 50.0\% to 90.0\% in agentic mode, while precision decreases slightly from 62.5\% to 60.0\%. This suggests that repository evidence makes GPT-5.6 more willing to assign \textsc{Contextual\_Risk} when it identifies a plausible attacker-reachable path.

On the two majority classes, \textsc{Hardening\_Recommendation} and \textsc{False\_Positive}, agentic exploration improves both precision and recall together for Claude Sonnet~5 (F1-scores rising from 76.6\% to 91.2\% and from 82.3\% to 95.0\%, respectively), indicating that additional evidence sharpens its decisions on both classes simultaneously rather than trading one off for the other. GPT-5.6 instead shows a precision-recall trade-off within \textsc{Hardening\_Recommendation}: agentic mode raises precision sharply (62.7\% to 94.7\%) while recall falls (92.2\% to 69.6\%), as GPT-5.6 becomes more selective about applying this label once repository evidence is available, while its \textsc{False\_Positive} F1-score still improves substantially (74.8\% to 90.6\%).

GLM-5.2's class-wise results are consistent with the reasoning-label mismatch identified in the overall analysis. Its \textsc{False\_Positive} performance moves in the opposite direction from the other two models under agentic exploration: both precision (77.3\% to 72.3\%) and recall (54.4\% to 43.5\%) decline, rather than improving as they do for Claude Sonnet~5 and GPT-5.6. This decline is mirrored by a small drop in \textsc{Hardening\_Recommendation} precision (52.0\% to 47.9\%) despite comparable recall, suggesting that additional agentic evidence is, if anything, pulling genuine false positives into the \textsc{Hardening\_Recommendation} class rather than resolving them correctly -- the same mismatch between GLM-5.2's generated reasoning and its selected label described above, now visible directly in the per-class numbers rather than only in individual examples.

\begin{tcolorbox}[colback=gray!8, colframe=black, boxrule=0.5pt,
                   title=\textbf{RQ2 Takeaway}, fonttitle=\bfseries,
                   left=2mm, right=2mm, top=1mm, bottom=1mm]
Context-aware analysis of security findings can be automated at scale when the framework grounds its decisions in evidence collected directly from the artifact repository. SAFE shows that repository-level exploration can substantially improve classification quality over zero-shot contextual analysis for models that effectively use the available tools.
\end{tcolorbox}

\subsection{Cost Analysis}
\label{subsec:cost_analysis}
We analyze the cost of applying SAFE for artifact evaluation. The primary cost comes from LLM-based reasoning, while static analysis using Semgrep and Trivy adds negligible overhead.
Table~\ref{tab:safe_cost} reports the monetary and computational cost of evaluating 250 findings across 1,500 runs (three models and two analysis modes). Cost per finding varies by more than an order of magnitude across models, from \$0.022 for GLM-5.2 in zero-shot mode to \$0.229 for Claude Sonnet~5 in agentic mode. Much of this spread reflects per-token pricing rather than differences in how much a model reads: GLM-5.2's zero-shot token consumption (21{,}272 tokens/finding) is close to GPT-5.6's (20{,}283 tokens/finding), yet GLM-5.2 costs roughly 6.5$\times$ less per finding, which shows that GLM-5.2's lower cost is primarily a pricing effect rather than a token-efficiency gain.
\begin{table}[htbp]
    \centering
    \caption{Cost of running SAFE across 1,500 runs.}
    \label{tab:safe_cost}
    \vspace{-1mm}
    \begin{tabularx}{\linewidth}{XlXXp{1.8cm}}
        \toprule
        \textbf{Mode} & \textbf{Model} &
        \textbf{Total Cost} & \textbf{Cost/
        Finding} & \textbf{Avg. Tokens/ Finding} \\
        \midrule
        \multirow{3}{*}{Zero-shot}
        &  Sonnet 5 & \$23.24 & \$0.093 & 33{,}331 \\
        & GPT-5.6         & \$36.06 & \$0.144 & 20{,}283 \\
        & GLM-5.2         & \$5.57  & \$0.022 & 21{,}272 \\
        \midrule
        \multirow{3}{*}{Agentic}
        &  Sonnet 5 & \$57.35 & \$0.229 & 99{,}870 \\
        & GPT-5.6         & \$46.53 & \$0.186 & 61{,}492 \\
        & GLM-5.2         & \$6.62  & \$0.026 & 23{,}482 \\
        \bottomrule
    \end{tabularx}
    \vspace{-2mm}
\end{table}

Agentic mode increases cost for every model, but by very different margins, and the size of the increase tracks the accuracy gain reported in Section~\ref{subsubsec:overall_analysis_result}. Claude Sonnet~5's cost per finding rises 2.5$\times$ from zero-shot to agentic mode (\$0.093 to \$0.229), alongside its largest accuracy gain of the three models; GPT-5.6's cost rises a more modest 1.3$\times$ (\$0.144 to \$0.186), alongside a smaller but still substantial accuracy gain. GLM-5.2's cost, by contrast, rises only 1.2$\times$ (\$0.022 to \$0.026) -- essentially the price of one extra API round-trip -- and its accuracy does not improve.

\subsection{Failure Analysis}
We examine cases where the model does not produce a usable structured prediction. We distinguish between (i) a \emph{schema-repair event}, where the initial response fails schema validation but a repair request produces a valid prediction, and (ii) a \emph{permanent failure}, where no valid prediction is produced after the repair attempts.
\begin{table}[htbp]
    \centering
    \caption{Failure analysis across the runs.}
    \label{tab:safe_failure_taxonomy}
    \begin{tabularx}{\linewidth}{llcc}
        \toprule
        \textbf{Mode} & \textbf{Model} & \textbf{Schema-Repair} & 
        \textbf{Permanent Failures} \\
        \midrule
        \multirow{3}{*}{Zero-shot}
        & GPT-5.6         & 0 & 0 \\
        & GLM-5.2         & 7 & 0 \\
        &  Sonnet 5 & 1 & 2 \\
        \midrule
        \multirow{3}{*}{Agentic}
        & GPT-5.6         & 0 & 0 \\
        & GLM-5.2         & 0 & 0 \\
        &  Sonnet 5 & 0 & 1 \\
        \bottomrule
    \end{tabularx}
\end{table}
Schema repair is available in both analysis modes, although all observed repair events occurred in zero-shot runs. As shown in Table~\ref{tab:safe_failure_taxonomy}, GLM-5.2 required schema repair for 7 of 250 zero-shot findings and Sonnet~5 for 1, while GPT-5.6 required none. Permanent output failures were rare, with two zero-shot failures and one agentic failure for Sonnet~5. Thus, the near-zero values in Table~\ref{tab:safe_failure_taxonomy} indicate high structured-output reliability, rather than error-free classification.
Table~\ref{tab:safe_permanent_failures} in Appendix lists all permanent failures.



\section{Discussion}

RQ1 shows that scanner rules alone are not enough to determine the real security relevance of a finding. The same unsafe construct can represent a \texttt{CONTEXTUAL\_RISK} in one artifact and a \texttt{HARDENING\_RECOMMENDATION} in another. This distinction depends on how the construct is used and therefore requires artifact-specific context. Manual review can provide this context, but it does not scale to the large number of findings produced across the artifact corpus. RQ2 shows that SAFE can automate much of this contextual assessment. In agentic mode, Claude Sonnet~5 and GPT-5.6 improve binary F1-score by 13.1 and 7.6 percentage points, respectively, and Claude Sonnet~5 reaches 94.4\% accuracy. The results also show that the benefit depends on the underlying model. Claude Sonnet~5 and GPT-5.6 make effective use of repository evidence, while GLM-5.2 benefits less from the same process (Section~\ref{appxsec:tool_usage} in Appendix). This suggests that reliable use of repository evidence is an important factor when selecting models for context-aware security analysis.

\subsection{Implications for Artifact Evaluation Practice}

We aim to further refine the efficiency and usability of SAFE to facilitate its adoption within artifact evaluation workflows. Prior work has also explored tool-assisted vulnerability assessment, showing how decision-support tools can aid security analysis~\cite{zhang2026tool}. We encourage authors to use SAFE to document execution assumptions, minimize unsafe practices, and ensure that artifacts do not unintentionally expose exploitable behavior when reused. In addition, we encourage artifact reviewers to move beyond purely functional validation and consider basic security assessment, particularly when assessing whether flagged issues are realistically exploitable. We also encourage artifact evaluation committee chairs and top-tier venues to promote the use of lightweight security assessment frameworks like SAFE, to enable a broader shift toward security-aware evaluation practices. Prior longitudinal evidence shows that conference artifact policies can influence artifact-sharing practices, particularly the adoption of stable hosting services~\cite{vansteenhuyse2026not}, suggesting that AE guidelines can similarly encourage more security-conscious artifact release.

We advocate for introducing optional security checks or badges as part of artifact evaluation, supported by automated tools and simple guidelines. At its current stage, SAFE is not intended to serve as a mandatory evaluation criterion; rather, it can be adopted as an optional security assessment layer to encourage responsible artifact release.

This proposal extends an established practice rather than introducing a new one: AE venues already award badges for availability, functionality, and reproducibility without requiring every artifact to meet every criterion. A security-considered badge can follow the same model, recognizing that an artifact has been assessed with a framework like SAFE and that any flagged \texttt{CONTEXTUAL\_RISK} findings have been documented or addressed, giving reviewers and downstream reusers a visible signal that security was considered, while keeping the barrier to adoption low.


\textit{Artifact Security Checklist.} To support practical adoption, we provide a lightweight checklist that can be used by authors, and AE chairs and reviewers, during artifact evaluation:
\begin{itemize}
    \item Does the artifact process external or user-controlled input?
    \item Are unsafe operations like deserialization and shell execution properly validated or restricted?
    \item Are third-party dependencies free from known critical vulnerabilities?
    \item Is the intended execution context, such as offline experiment vs. deployment, clearly documented?
    \item Are assumptions about trust boundaries and input sources explicitly stated?
\end{itemize}
This checklist complements automated analysis by encouraging explicit consideration of security-relevant aspects that are often implicit in research code.

\subsection{Case Studies}
To qualitatively evaluate SAFE's reasoning, we present two case studies drawn from its agentic evaluation runs. These are SAFE predictions: each listing below reproduces the model's actual tool-call trajectory and its actual structured output, unedited apart from truncation for space. 

\subsubsection{Case Study 1: 
}

Semgrep flagged an unpinned GitHub Actions reference in \texttt{.github/workflows/build.yml:460} of a proof-of-concept artifact for a CI cache-leakage vulnerability. The step \texttt{uses: actions/upload-artifact@v3} resolves a mutable tag rather than a pinned 40-character commit SHA (CWE-1357, CWE-353), the same class of CI supply-chain weakness discussed qualitatively in Section~\ref{subsec:represent_cases}, Case~2. In zero-shot mode, GPT-5.6 identifies the unpinned reference but, reasoning from the evidence packet alone, cannot confirm whether the action executes after sensitive material has entered the workspace and classifies it as \texttt{HARDENING\_RECOMMENDATION}. In agentic mode, the same model instead reads the surrounding workflow and the artifact's README before answering.



\begin{lstlisting}[
caption={Case Study 1: GPT-5.6 agentic trajectory (A078)},
label={lst:casestudy1_traj},
basicstyle=\ttfamily,
columns=fullflexible,
keepspaces=true,
breaklines=true,
showstringspaces=false,
frame=single,
framesep=4pt,
xleftmargin=4pt,
xrightmargin=4pt
]
Step 1: read_file(build.yml, lines 1-90)
        read_file(build.yml, lines 330-470)
        read_file(README.md, lines 1-30)
Step 2: final classification emitted
        (3 tool calls, 2 agent steps)
\end{lstlisting}

Reading lines 330-470 reveals that the same CI job decodes Apple and Windows code-signing keys into workspace files (\texttt{build.yml:375-392}) \emph{before} the mutable action executes, and the README confirms the workflow is triggered on a documented \texttt{github-action-*} branch. This evidence predicted the classification as \textsc{CONTEXTUAL\_RISK} (output is shown in the Appendix Listing~\ref{lst:casestudy1_output}). This matches the ground truth exactly, whose justification independently identifies the same mechanism: ``the action is resolved across a live external trust boundary, and its code runs after sensitive files have been created in the workspace.'' 

\subsubsection{Case Study 2:
}

A Trivy-reported vulnerability (CWE-409) in an artifact, which is a self-replicating prompt-injection worm simulation built on LLaVA, concerns a decompression-bomb bypass in \texttt{urllib3} 2.2.1 during HTTP-redirect handling. The dependency file is UTF-16 encoded, which defeats SAFE's automated package-usage check and reports the package as undeclared. This case shows that exploration volume alone does not determine whether a model resolves such a finding correctly.

\begin{lstlisting}[
caption={Case Study 2: Sonnet 5 agentic trajectory (A088)},
label={lst:casestudy2_traj},
basicstyle=\ttfamily,
columns=fullflexible,
keepspaces=true,
breaklines=true,
showstringspaces=false,
frame=single,
framesep=4pt,
xleftmargin=4pt,
xrightmargin=4pt
]
Step 1: search_repository("urllib3");
        read_file(requirements.txt, 1-5)
Step 2: search_repository("aiohttp");
        search_repository("urllib")
Step 3: read_file(requirements.txt, 1-200)
Step 4: read_file(requirements.txt, 200-400)
Step 5: search_repository("import urllib3");
        search_repository("preload_content");
        inspect_package_usage("urllib3")
Step 6: read_file(Run_LLaVa.py, 1-100)
Step 7: search_repository("load_image")
Step 8: final classification emitted
        (11 tool calls, 8 agent steps)
\end{lstlisting}

When the automated package check returns a negative result, Claude Sonnet~5 does not accept it: it falls back to reading the raw manifest, confirms \texttt{urllib3==2.2.1} is genuinely pinned, and searches specifically for \texttt{load\_image}, locating \texttt{Run\_LLaVa.py:75--81}, where an image URL is fetched with \texttt{requests.get()} without restriction.

GPT-5.6, given the identical finding, uses \emph{more} tool calls (15 across 4 agent steps) but reaches \texttt{HARDENING\_RECOMMENDATION}. Its exploration is not shallow as it locates three genuine streaming-request call sites in the artifact's LLaVA serving infrastructure (\texttt{gradio\_web\_server.py}, \texttt{controller.py}, \texttt{test\_message.py}) and correctly confirms the vulnerable dependency is present and reachable. What it does not locate is \texttt{Run\_LLaVa.py}'s \texttt{load\_image()}, and so it reasons about a weaker, internally-routed attack surface (a controller defaulting to \texttt{localhost}) rather than the artifact's documented externally-facing one, assessing attacker control as uncertain. Both models gather substantial, genuine evidence; only one gathers the evidence that resolves the finding, showing that the agentic accuracy gains reported in Section~\ref{subsubsec:overall_analysis_result} depend on \emph{which} evidence a model's search strategy surfaces, not merely on how much evidence it gathers.

\vspace{-1.3mm}
\subsection{Limitations and Future Work}

SAFE currently relies on Semgrep and Trivy for candidate findings, so it inherits the limitations of static analysis and cannot detect risks outside their coverage. Its evidence-gathering tools are also read-only and do not execute artifacts. As a result, runtime-dependent conditions, such as environment variables or command-line inputs, must be inferred from code and documentation. In addition, our evaluation covers four security venues over three years, so the generality of the taxonomy and performance trends beyond this setting remains to be studied.
Future work can extend SAFE in three directions: adding more scanners to broaden candidate coverage, introducing lightweight sandboxed dynamic analysis for findings that remain \texttt{uncertain} after static inspection, and evaluating SAFE on artifacts from a wider range of venues and research domains.
\section{Conclusion}
\label{sec:conclusion}

Security remains a largely overlooked dimension in current artifact evaluation processes, which primarily focus on functional correctness and reproducibility. While these aspects are essential, they do not capture whether the released artifacts may introduce unintended security risks when reused, extended, or deployed.
In this work, we demonstrate the need for context-aware security analysis in research artifacts. Our findings demonstrate that the security relevance of a finding depends on execution context, attacker control, and intended usage. Through SAFE, we show that incorporating structured reasoning enables meaningful distinctions between benign experimental patterns and realistic security risks.
We position SAFE as a first step toward integrating security into artifact evaluation in a practical and lightweight manner. SAFE provides a framework that supports authors, reviewers, and artifact evaluation committees in identifying and reasoning about potential risks. The proposed approach opens the direction for future systems that combine automated analysis with deeper contextual understanding, ultimately contributing to more secure and responsible release of research artifacts. This shift is essential for aligning artifact evaluation with the growing need for secure, reusable, and trustworthy research outputs.

\bibliographystyle{ACM-Reference-Format}
\bibliography{refs}

@inproceedings{olszewski2023get,
  title={" Get in Researchers; We're Measuring Reproducibility": A Reproducibility Study of Machine Learning Papers in Tier 1 Security Conferences},
  author={Olszewski, Daniel and Lu, Allison and Stillman, Carson and Warren, Kevin and Kitroser, Cole and Pascual, Alejandro and Ukirde, Divyajyoti and Butler, Kevin and Traynor, Patrick},
  booktitle={Proceedings of the 2023 ACM SIGSAC conference on computer and communications security},
  pages={3433--3459},
  year={2023}
}

@inproceedings{winter2022retrospective,
  title={A retrospective study of one decade of artifact evaluations},
  author={Winter, Stefan and Timperley, Christopher S and Hermann, Ben and Cito, J{\"u}rgen and Bell, Jonathan and Hilton, Michael and Beyer, Dirk},
  booktitle={Proceedings of the 30th ACM joint European software engineering conference and symposium on the foundations of software engineering},
  pages={145--156},
  year={2022}
}

@inproceedings{guilloteau2024longevity,
  title={Longevity of artifacts in leading parallel and distributed systems conferences: A review of the state of the practice in 2023},
  author={Guilloteau, Quentin and Ciorba, Florina and Poquet, Millian and Goepp, Dorian and Richard, Olivier},
  booktitle={Proceedings of the 2nd ACM Conference on Reproducibility and Replicability},
  pages={121--133},
  year={2024}
}

@article{guevara2024research,
  title={Research artifacts for human-oriented experiments in software engineering: An ACM badges-driven structure proposal},
  author={Guevara-Vega, Cathy and Bern{\'a}rdez, Beatriz and Cruz, Margarita and Dur{\'a}n, Amador and Ruiz-Cort{\'e}s, Antonio and Solari, Martin},
  journal={Journal of Systems and Software},
  volume={218},
  pages={112187},
  year={2024},
  publisher={Elsevier}
}

@article{beyer2025reproducibility,
  title={Reproducibility and replication of research results: A special issue for RRRR 2022},
  author={Beyer, Dirk and Hartmanns, Arnd},
  journal={International Journal on Software Tools for Technology Transfer},
  volume={27},
  number={4},
  pages={397--401},
  year={2025},
  publisher={Springer}
}

@article{baek2026artisan,
  title={Artisan: Agentic Artifact Evaluation},
  author={Baek, Doehyun and Pradel, Michael},
  journal={arXiv preprint arXiv:2602.10046},
  year={2026}
}

@article{wu2026agent,
  title={Agent-Based Software Artifact Evaluation},
  author={Wu, Zhaonan and Zhao, Yanjie and Chen, Zhenpeng and Wang, Zheng and Wang, Haoyu},
  journal={arXiv preprint arXiv:2602.02235},
  year={2026}
}

@article{hermann2022has,
  author={Hermann, Ben},
  journal={IEEE Security \& Privacy}, 
  title={What Has Artifact Evaluation Ever Done for Us?}, 
  year={2022},
  volume={20},
  number={5},
  pages={96-99},
  doi={10.1109/MSEC.2022.3184234}}

@inproceedings{beyer2025artifact,
  title={Artifact Evaluations for Stronger Research Results},
  author={Beyer, Dirk and Winter, Stefan},
  booktitle={Proceedings of the 33rd ACM International Conference on the Foundations of Software Engineering},
  pages={1234--1237},
  year={2025}
}

@inproceedings{barr2023continuously,
  title={Continuously Accelerating Research},
  author={Barr, Earl and Bell, Jonathan and Hilton, Michael and Mechtaev, Sergey and Timperley, Christopher},
  booktitle={2023 IEEE/ACM 45th International Conference on Software Engineering: New Ideas and Emerging Results (ICSE-NIER)},
  pages={123--128},
  year={2023},
  organization={IEEE}
}

@inproceedings{heye2025supporting,
  title={Supporting Artifact Evaluation with LLMs: A Study with Published Security Research Papers},
  author={Heye, David and Kindermann, Karl and Decker, Robin and Lohm{\"o}ller, Johannes and Belova, Anastasiia and Geisler, Sandra and Wehrle, Klaus and Pennekamp, Jan},
  booktitle={2025 IEEE International Conference on Big Data (BigData)},
  pages={5077--5085},
  year={2025},
  organization={IEEE}
}

@inproceedings{yao2022react,
  title={React: Synergizing reasoning and acting in language models},
  author={Yao, Shunyu and Zhao, Jeffrey and Yu, Dian and Du, Nan and Shafran, Izhak and Narasimhan, Karthik R and Cao, Yuan},
  year={2022}
}

@misc{sedghpour2024artifact,
      title={Artifact Evaluation for Distributed Systems: Current Practices and Beyond}, 
      author={Mohammad Reza Saleh Sedghpour and Alessandro Vittorio Papadopoulos and Cristian Klein and Johan Tordsson},
      year={2024},
      eprint={2406.13045},
      archivePrefix={arXiv},
      primaryClass={cs.DC},
      url={https://arxiv.org/abs/2406.13045}, 
}

@inproceedings{malik2020artifact,
  title={Artifact description/artifact evaluation: A reproducibility bane or a boon},
  author={Malik, Tanu},
  booktitle={Proceedings of the 4th International Workshop on Practical Reproducible Evaluation of Computer Systems},
  pages={1--1},
  year={2020}
}

@inproceedings{olszewski2025reproducibility,
  title={Reproducibility in Applied Security Conferences: An 11-Year Review on Artifacts and Evaluation Committees},
  author={Olszewski, Daniel and Lu, Allison and Crowder, Anna and Bennett, Nathaniel and Layton, Seth and Varma Bhupathiraju, Sri Hrushikesh and Tucker, Tyler and Kalgutkar, Siddhant and Ver Helst, Hunter and Stillman, Carson and others},
  booktitle={Proceedings of the 3rd ACM Conference on Reproducibility and Replicability},
  pages={96--107},
  year={2025}
}

@inproceedings{guilloteau2024artifact,
  TITLE = {{Artifact Evaluations as Authors and Reviewers : Lessons, Questions, and Frustrations}},
  AUTHOR = {Guilloteau, Quentin and Poquet, Millian and M{\"u}ller Kornd{\"o}rfer, Jonas H and Ciorba, Florina M},
  URL = {https://hal.science/hal-04764265},
  BOOKTITLE = {{Community Workshop on Practical Reproducibility in HPC}},
  ADDRESS = {Atlanta (GA), United States},
  YEAR = {2024},
  MONTH = Nov,
  HAL_ID = {hal-04764265},
  HAL_VERSION = {v1},
}

@article{muttakin2026state,
  title={The State of Open Science in Software Engineering Research: A Case Study of ICSE Artifacts},
  author={Muttakin, Al and Mondal, Saikat and Roy, Chanchal K},
  journal={arXiv preprint arXiv:2601.02066},
  year={2026}
}

@inproceedings {olszewski2025sok,
author = {Daniel Olszewski and Tyler Tucker and Kevin R. B. Butler and Patrick Traynor},
title = {{SoK}: Towards a Unified Approach to Applied Replicability for Computer Security},
booktitle = {34th USENIX Security Symposium (USENIX Security 25)},
year = {2025},
isbn = {978-1-939133-52-6},
address = {Seattle, WA},
pages = {469--488},
url = {https://www.usenix.org/conference/usenixsecurity25/presentation/olszewski},
publisher = {USENIX Association},
month = aug
}

@article{zilberman2020thoughts,
author = {Zilberman, Noa and Moore, Andrew W.},
title = {Thoughts about Artifact Badging},
year = {2020},
issue_date = {April 2020},
publisher = {Association for Computing Machinery},
address = {New York, NY, USA},
volume = {50},
number = {2},
issn = {0146-4833},
journal = {SIGCOMM Comput. Commun. Rev.},
month = may,
pages = {60–63},
numpages = {4}
}

@article{heumuller2020publish,
  title={Publish or perish, but do not forget your software artifacts},
  author={Heum{\"u}ller, Robert and Nielebock, Sebastian and Kr{\"u}ger, Jacob and Ortmeier, Frank},
  journal={Empirical Software Engineering},
  volume={25},
  number={6},
  pages={4585--4616},
  year={2020},
  publisher={Springer}
}

@article{guo2023mitigating,
  title={Mitigating false positive static analysis warnings: Progress, challenges, and opportunities},
  author={Guo, Zhaoqiang and Tan, Tingting and Liu, Shiran and Liu, Xutong and Lai, Wei and Yang, Yibiao and Li, Yanhui and Chen, Lin and Dong, Wei and Zhou, Yuming},
  journal={IEEE Transactions on Software Engineering},
  volume={49},
  number={12},
  pages={5154--5188},
  year={2023},
  publisher={IEEE}
}

@inproceedings{kang2022detecting,
  title={Detecting false alarms from automatic static analysis tools: How far are we?},
  author={Kang, Hong Jin and Aw, Khai Loong and Lo, David},
  booktitle={Proceedings of the 44th International Conference on Software Engineering},
  pages={698--709},
  year={2022}
}

@misc{trivy,
  author       = {{Aqua Security}},
  title        = {Trivy: Vulnerability Scanner for Containers and Other Artifacts},
  year         = {2026},
  url          = {https://github.com/aquasecurity/trivy},
  note         = {{Accessed: 2026-08-13}}
}

@misc{semgrep,
  author       = {{Semgrep, Inc.}},
  title        = {Semgrep: Static Analysis Tool for Code Security},
  year         = {2026},
  url          = {https://github.com/semgrep/semgrep},
  note         = {{Accessed: 2026-08-13}}
}

@inproceedings{vansteenhuyse2026not,
  title={Not All Those Who Share Are Lost: Analyzing 25 Years of Cybersecurity Artifact Sharing Practices Through Automated Discovery},
  author={Vansteenhuyse, Daan and Bols, Arthur and Desmet, Lieven and Le Pochat, Victor and Van Bulck, Jo and Bognar, Marton},
  booktitle={Proceedings of the 35th USENIX Security Symposium},
  year={2026}
}

@misc{reprodb2026,
  title        = {ReproDB: Research Artifact and Artifact Evaluation Database},
  author       = {{ReproDB}},
  year         = {2026},
  howpublished = {\url{https://reprodb.github.io/}},
  note         = {Accessed: 2026-08-09}
}

@misc{usenixsecurity26,
  author       = {{USENIX Association}},
  title        = {{USENIX Security '26 Call for Papers}},
  year         = {2026},
  howpublished = {\url{https://www.usenix.org/conference/usenixsecurity26/call-for-papers}},
  note         = {Accessed: 2026-08-21}
}

@inproceedings {ghosh2025wasn,
author = {Rajdeep Ghosh and Shiladitya De and Mainack Mondal},
title = {"I wasn{\textquoteright}t sure if this is indeed a security risk": Data-driven Understanding of Security Issue Reporting in {GitHub} Repositories of Open Source npm Packages},
booktitle = {34th USENIX Security Symposium (USENIX Security 25)},
year = {2025},
isbn = {978-1-939133-52-6},
address = {Seattle, WA},
pages = {2145--2164},
url = {https://www.usenix.org/conference/usenixsecurity25/presentation/ghosh},
publisher = {USENIX Association},
month = aug
}

@article{ye2024detecting,
  title={Detecting command injection vulnerabilities in Linux-based embedded firmware with LLM-based taint analysis of library functions},
  author={Ye, Junjian and Fei, Xincheng and de Carnavalet, Xavier de Carn{\'e} and Zhao, Lianying and Wu, Lifa and Zhang, Mengyuan},
  journal={Computers \& Security},
  volume={144},
  pages={103971},
  year={2024},
  publisher={Elsevier}
}

@inproceedings{oh2024poisoned,
  title={Poisoned chatgpt finds work for idle hands: Exploring developers’ coding practices with insecure suggestions from poisoned ai models},
  author={Oh, Sanghak and Lee, Kiho and Park, Seonhye and Kim, Doowon and Kim, Hyoungshick},
  booktitle={2024 IEEE Symposium on Security and Privacy (SP)},
  pages={1141--1159},
  year={2024},
  organization={IEEE}
}

@inproceedings{zhang2026tool,
  title={Tool-Assisted CVSS Vulnerability Scoring: A Controlled Quantitative Study of Human Assessment},
  author={Zhang, Siqi and Cai, Minjie and Zhao, Lianying and de Carn{\'e} de Carnavalet, Xavier and Massacci, Fabio and Zhang, Mengyuan},
  booktitle={Proceedings of the 2026 CHI Conference on Human Factors in Computing Systems},
  pages={1--24},
  year={2026}
}

@inproceedings{robinson2026original,
  title={Original Sin of npm: A Study on Vulnerability Propagation in JavaScript Dependency Networks},
  author={Robinson, Michael and Halder, Sajal and Ahmed, Muhammad Ejaz and Ikram, Muhammad and Camtepe, Seyit and Kim, Hyoungshick},
  booktitle={Proceedings of the ACM Asia Conference on Computer and Communications Security},
  pages={1213--1227},
  year={2026}
}

@String{Computing = "Computing" }

@String{Computer = "{IEEE} Computer" }

@String{Springer = "Springer-Verlag" }

\appendix


\section{Ethical Considerations}

All artifacts evaluated in this study are drawn from already-published, publicly archived proceedings (CCS, NDSS, S\&P, and USENIX Security, 2023--2025); no embargoed or under-review material was analyzed, and classification was performed solely for the research evaluation reported in this paper. Our analysis did not actively exploit identified weaknesses or interact with associated live systems or services. SAFE performs static analysis and read-only repository inspection, and we report only the technical details necessary to explain the identified risks. We also avoid interpreting the presence of an insecure construct as evidence of negligence by artifact authors, since its practical relevance depends on execution context and realistic attacker influence.

We further emphasize that SAFE is intended to augment, rather than replace, expert security review. Its precision on \texttt{CONTEXTUAL\_RISK} is high, but its recall is deliberately conservative, so an artifact SAFE does not flag should not be interpreted as certified free of risk. Any security badge built on SAFE should state this distinction explicitly, signaling that an artifact underwent SAFE-assisted assessment rather than guaranteeing the absence of vulnerabilities.

\begin{table*}[htbp]
\centering
\caption{Exact download-failure reasons.}
\label{tab:download_failure_reasons}
\begin{tabularx}{\textwidth}{p{3.8cm} r X}
\toprule
\textbf{Failure reason} & \textbf{Count} & \textbf{Explanation} \\
\midrule
DNS/name-resolution failure
& 9
& The host name could not be resolved. Eight reported a temporary name-resolution failure and one reported ``Name or service not known.'' \\

GitHub repository not found/authentication failure
& 3
& Git returned ``Repository not found'' followed by authentication failure. This includes one GitHub topic page incorrectly treated as a repository target. \\

HTTP 403 Forbidden
& 3
& The server received the request but refused access. Two were Tor Project GitLab webpages and one was an Intel webpage. \\

HTTP 404 Not Found
& 2
& The URL resolved to a server, but the requested resource did not exist. \\

HTTP 410 Gone
& 2
& The Zenodo records were reported as permanently unavailable. \\

TLS/certificate failure
& 2
& One endpoint failed certificate-chain verification; another failed during TLS server-name negotiation. \\

GitLab authentication denied
& 1
& GitLab rejected repository access with an HTTP Basic/token authentication error. \\

HTTP 503 Service Unavailable
& 1
& The server was temporarily unable to handle the request. \\

Read timeout
& 1
& The server did not complete the response within the configured timeout. \\

\midrule
\textbf{Total}
& \textbf{24}
& \\
\bottomrule
\end{tabularx}
\end{table*}

\section{Artifact Download Failures}
\label{appxsec:download_failure}
During artifact collection, a small number of artifacts could not be downloaded successfully. Table~\ref{tab:download_failure_reasons} summarizes the exact causes of these failures. In total, 24 download attempts failed due to inaccessible or unavailable resources, including DNS resolution errors, repository access or authentication failures, HTTP errors, TLS/certificate issues, and network timeouts. DNS and name-resolution errors were the most common cause, accounting for 9 of the 24 failures. These artifacts were excluded from subsequent analysis because their contents could not be retrieved.

\section{Artifact Collection and Exclusion Details}
\label{app:artifact_collection_details}
Table~\ref{tab:complete_exclusion_reasons} summarizes the reasons for excluding successfully collected artifacts that were not suitable for subsequent code-based security analysis. Of the 1,892 successfully collected artifacts, 504 were excluded because they contained websites, empty directories, datasets, documentation, model checkpoints, formal specifications, incomplete repository content, or other material without suitable source code. This resulted in a final cohort of 1,388 code-bearing artifacts used in the subsequent analysis.

\begin{table*}[t]
\centering
\caption{Complete exclusion reasons.}
\label{tab:complete_exclusion_reasons}
\begin{tabularx}{\textwidth}{p{5.3cm} r X}
\toprule
\textbf{Exclusion or hold-out reason} & \textbf{Count} & \textbf{Explanation} \\
\midrule

Website or HTML snapshot
& 327
& The downloaded material was a project page, publication page, or HTML snapshot rather than a source-code collection. \\

No local content
& 12
& Acquisition was marked successful, but the resulting artifact directory contained no usable files. \\

Dataset without recognized code
& 10
& Data files were present, but no recognized implementation code was detected. \\

Documentation without recognized code
& 10
& The artifact contained reports, instructions, or other documentation but no recognized implementation code. \\

Mixed non-source artifact
& 2
& Multiple non-source file types were present without recognized code. \\

Model/checkpoint without recognized code
& 1
& The artifact contained model-weight or checkpoint material without implementation code. \\

No code found during automated secondary inspection
& 127
& Ambiguous artifacts were inspected more deeply, but no recognized code, notebook code cells, or executable scripts were found. \\

Build/package metadata only
& 5
& Build or dependency metadata existed, but no implementation files were available. These were conservatively held out. \\

Formal specifications requiring specialized tooling
& 5
& Formal specification files were present, but they were not suitable for the planned Semgrep/Trivy workflow. \\

Missing Git LFS/submodule content
& 5
& Repository metadata referenced content that had not been retrieved, so code availability could not be established. \\

\midrule
\textbf{Total}
& \textbf{504}
& \\
\bottomrule
\end{tabularx}
\end{table*}

\section{Normalized Finding Schema}
\label{app:finding_schema}

Table~\ref{tab:finding_schema} summarizes the common representation used to
combine Semgrep code findings, Trivy dependency vulnerabilities, and Trivy
secret findings. Tool-specific fields that are not applicable to a particular
finding type are left empty.

\begin{table*}[t]
\centering
\caption{Normalized representation of candidate security findings.}
\label{tab:finding_schema}
\begin{tabularx}{\textwidth}{l p{3.9cm} X}
\toprule
\textbf{Group} & \textbf{Fields} & \textbf{Description} \\
\midrule

Artifact identity &
\texttt{conference}, \texttt{artifact\_id},
\texttt{artifact\_key} &
Identifies the conference-year collection and artifact. The composite
\texttt{artifact\_key} provides the globally unique identifier. \\

Finding provenance &
\texttt{tool}, \texttt{finding\_id},
\texttt{category} &
Identifies the originating scanner, its rule or advisory identifier, and the
normalized finding class: code issue, dependency vulnerability, or potential
secret. \\

Severity &
\texttt{severity\_raw}, \texttt{cvss} &
Stores the scanner-reported severity and the highest available CVSS score.
The raw severity is retained because the tools use different severity
semantics. \\

Location &
\texttt{file}, \texttt{start\_line},
\texttt{start\_column}, \texttt{end\_line},
\texttt{end\_column} &
Records the normalized artifact-relative file path and available source
position. \\

Description &
\texttt{message} &
Contains the scanner-provided title or description of the candidate issue. \\

Dependency metadata &
\texttt{package}, \texttt{version},
\texttt{fixed\_version} &
Records the affected package, installed version, and available fixed version
for dependency findings. \\

Weakness metadata &
\texttt{cwe} &
Contains associated CWE identifiers when supplied by the scanner or
vulnerability metadata. CVE or advisory identifiers are represented by
\texttt{finding\_id}. \\

Audit metadata &
\texttt{file\_raw},
\texttt{raw\_duplicate\_count} &
Preserves the original scanner path and the number of identical raw scanner
records represented by a normalized finding. \\

\bottomrule
\end{tabularx}
\end{table*}



\section{Contextual-Assessment Sampling Procedure}
\label{app:contextual_sampling}

We construct the contextual-assessment sample from the deduplicated
finding table. The sampling unit is an individual candidate finding,
while the stratification unit is a tool-qualified finding type,
defined as $(\texttt{tool},\texttt{finding\_id})$. This distinction is
necessary because a Semgrep rule and a Trivy vulnerability identifier
represent different classes of scanner output.

For every finding type, we compute artifact-level prevalence as

\[
P(f) =
\left|
\left\{
a \mid \text{artifact } a \text{ contains at least one instance of } f
\right\}
\right|.
\]

Multiple occurrences of the same type within one artifact therefore
contribute only once to its prevalence. This prevents large or repetitive
repositories from disproportionately determining which types are selected.

We retain finding types for which $P(f) \geq 30$. This results in 287
eligible types from the 4,194 tool-qualified types observed in the
deduplicated finding table. We rank these types in descending order of
artifact prevalence and retain the first 50. Ties are resolved
deterministically by scanner name followed by finding identifier. A tie
occurred at the selection boundary: the types ranked 50 and 51 both
occurred in 85 artifacts. The predetermined deterministic tie-breaking
rule selected the type at rank 50.

For every selected type, we choose five distinct artifacts without
replacement. We then choose one occurrence of that type from each
selected artifact. Selection is performed through an ordering derived
from SHA-256 over the fixed seed, scanner identifier, finding identifier,
artifact identifier, and, where necessary, a content-derived finding
identifier. This procedure provides pseudo-random selection while making
the result independent of input-row ordering and reproducible across
repeated executions.


Table~\ref{tab:contextual_sample} summarizes the successive filtering
and sampling stages. Starting from 132,431 deduplicated candidate
findings, we identified 4,194 tool-qualified finding types. Of these,
287 occurred in at least 30 distinct artifacts. Selecting the 50 most
artifact-prevalent types and sampling five instances per type produced
the final set of 250 candidates.

\begin{table}[htbp]
\centering
\caption{Construction of the contextual-assessment sample.}
\label{tab:contextual_sample}
\begin{tabular}{lr}
\toprule
\textbf{Sampling property} & \textbf{Count} \\
\midrule
Deduplicated candidate findings & 132,431 \\
Tool-qualified finding types & 4,194 \\
Types occurring in $\geq 30$ artifacts & 287 \\
Selected prevalent types & 50 \\
Instances sampled per type & 5 \\
Total sampled instances & 250 \\
Distinct sampled artifacts & 176 \\
\bottomrule
\end{tabular}
\end{table}

The sampled instances cover all twelve conference-year collections.
As shown in Table~\ref{tab:contextual_sample_distribution}, the sample
contains 57 instances from 2023, 83 from 2024, and 110 from 2025.
This distribution is reported descriptively because conference and year
were not used as sampling strata.

\begin{table}[htbp]
\centering
\caption{Conference-year distribution of the contextual-assessment sample.}
\label{tab:contextual_sample_distribution}
\begin{tabular}{lrrr}
\toprule
\textbf{Conference} & \textbf{2023} & \textbf{2024} & \textbf{2025} \\
\midrule
CCS             & 14 & 26 & 37 \\
NDSS            &  5 & 13 & 18 \\
IEEE S\&P       &  8 & 21 & 15 \\
USENIX Security & 30 & 23 & 40 \\
\midrule
Total           & 57 & 83 & 110 \\
\bottomrule
\end{tabular}
\end{table}
\FloatBarrier

\section{Additional Representative Case}
\label{appxsec:add_rep_case}
\paragraph{Case: Running a Container as Root.}
\textbf{What the flaw is.} A Dockerfile that never sets a \texttt{USER}
instruction leaves its container running as root by default. On its own,
this is a mild issue -- most containers never face anything that would
take advantage of it. It becomes serious specifically when the process
inside that container is already the kind of thing that handles content
it does not control (for example, a service that parses network input or
a browser that visits arbitrary websites): if that process is ever
tricked into running attacker code, root-by-default is what turns "a bug
in one component" into "full control of the container, with a
potential path back to the host."

\textbf{What the code does.} Both findings are Dockerfile finds where the
final image has no \texttt{USER} instruction, so its main process runs as
root.

\textbf{The risky version, step by step.}
\begin{enumerate}
  \item The container's documented job is to run a browser-driven
    crawler that visits external websites and renders whatever content it
    finds there -- visiting attacker-reachable, uncontrolled pages is not
    a misuse of the tool, it is the tool's stated purpose.
  \item The instructions for running this container also mount a
    writable directory from the host machine into the container, and give
    the container access to the host's display server.
  \item If a page the crawler visits contains something that exploits a
    bug in the browser or one of its extensions, that exploit gets to run
    as root inside the container, rather than as a low-privilege user.
  \item Because the container also has a writable path to the host
    filesystem, a root-level compromise inside the container is not fully
    contained -- it has a route back out to the host.
  \item Result: the missing \texttt{USER} instruction is not the only
    piece needed, but combined with everything the crawler is already
    exposed to, it turns a plausible browser bug into a much worse
    outcome than it would be in a properly sandboxed container.
\end{enumerate}

\textbf{The hardening-only version, step by step.}
\begin{enumerate}
  \item A second Dockerfile has the identical gap -- no \texttt{USER}
    instruction, so the image also runs as root by default.
  \item This image, however, belongs to deeply nested, third-party build
    infrastructure. Nothing in the artifact's documented workflow calls
    it, and nothing in the artifact routes external, attacker-reachable
    content through it.
  \item Result: the container still runs as root, but there is no
    documented scenario in which that container ever processes anything
    an outside party controls. There is no bug for the missing privilege
    drop to make worse.
\end{enumerate}
\begin{table*}[htbp]
\centering
\caption{Comparison of the additional case pairs along the same four questions.
}
\label{tab:appxcase_comparison}
\begin{tabularx}{\linewidth}{@{}llp{3cm}p{3.3cm}XX@{}}
\toprule
\textbf{Cases} & \textbf{Version} & \textbf{Who can supply the input?} & \textbf{Reached in normal use?} & \textbf{What has to go right?} & \textbf{What happens if it does?} \\
\midrule
\multirow{2}{*}{Container privilege}
 & Risky            & Anyone hosting a page the crawler visits       & Yes -- visiting sites is the tool's job          & A code-execution bug in the browser or an extension  & Root access inside a container with a host-writable path \\
 & Hardening   & No one -- nothing external reaches this image  & No -- image is unused in the documented workflow  & --- & --- \\
\bottomrule
\end{tabularx}
\end{table*}

\section{Tools Description}
\label{appxsec:tool_desc}
Table~\ref{tab:tools} outlines the 20 read-only tools SAFE's agent uses to analyze artifacts, organized into five categories: repository structure and discovery, file and code inspection, Python AST-based semantic analysis, dependency and vulnerability analysis, and paper-context grounding.
\begin{table*}[htbp]
\centering
\caption{Tools used by SAFE for context-aware artifact analysis}
\label{tab:tools}
\begin{tabularx}{\textwidth}{l p{2.6cm} X}
\toprule
\textbf{Tool} & \textbf{Purpose} & \textbf{Functionality} \\
\midrule
\multicolumn{3}{l}{\textit{Repository Structure and Discovery}} \\
\midrule
\texttt{list\_repository\_tree} & Repository structure enumeration & Lists repository-relative files discovered by the artifact profiler to locate entrypoints, documentation, implementations, tests, and configuration. \\
\texttt{inspect\_entrypoints} & Entry-point and exposure discovery & Returns detected entrypoints together with repository-wide input and exposure signals. \\
\texttt{inspect\_dependency\_files} & Dependency manifest retrieval & Returns bounded contents of dependency manifests (e.g., requirements.txt, pyproject.toml, package.json, go.mod, Cargo.toml, Maven/Gradle files). \\
\midrule
\multicolumn{3}{l}{\textit{File and Code Inspection}} \\
\midrule
\texttt{read\_file} & Full-file inspection & Reads a repository-relative text file with line numbers, for README, configuration, entrypoint, or exact code inspection. \\
\texttt{read\_snippet} & Localized code inspection & Reads numbered lines around a specific file location to examine a flagged code region together with its surrounding execution context. \\
\texttt{search\_repository} & Repository-wide text search & Searches literal text across repository source and documentation to locate symbols, inputs, vulnerable APIs, and documentation references. \\
\texttt{find\_symbol\_references} & Lexical reference search & Finds lexical references to a function, class, variable, API, or command across source and documentation. \\
\midrule
\multicolumn{3}{l}{\textit{Python AST-Based Semantic Analysis}} \\
\midrule
\texttt{inspect\_python\_scope} & Enclosing-scope lookup & Returns the Python function or class enclosing a specific line. \\
\texttt{inspect\_python\_function} & AST-based function inspection & Uses Python AST parsing to inspect the function enclosing a line, including its parameters, decorators, calls, sensitive calls, and return values. \\
\texttt{find\_python\_definitions} & Symbol definition search & Finds AST definitions of a Python function, class, or method name across the repository. \\
\texttt{find\_python\_callers} & Call-site discovery & Finds AST call sites for a Python symbol and reports the calling function, file path, line, and arguments. \\
\texttt{inspect\_python\_imports} & Import inspection & Inspects AST imports in a Python file, including aliases and relative imports. \\
\texttt{trace\_python\_variable} & Intra-function data-flow tracing & Collects intra-function AST evidence about a variable's parameters, prior assignments, calls, and possible input-origin hints. \\
\texttt{trace\_python\_reachability} & Reachability path construction & Builds bounded reverse Python call paths from a symbol toward detected entrypoints; results are name-based evidence rather than proof of reachability. \\
\texttt{find\_python\_security\_operations} & Security-sensitive operation discovery & Finds AST calls to security-sensitive Python operations across the repository, including command execution, deserialization, eval, network access, and archive extraction. \\
\midrule
\multicolumn{3}{l}{\textit{Dependency and Vulnerability Analysis}} \\
\midrule
\texttt{inspect\_package\_usage} & Package usage inspection & Inspects dependency declarations and Python imports for a distribution package; does not by itself prove CVE-feature reachability. \\
\texttt{inspect\_dependency\_graph} & Dependency graph lookup & Returns the checked-in dependency graph and the flagged package's inferred direct or transitive relationship to the artifact. \\
\texttt{inspect\_vulnerability\_advisory} & Advisory evidence retrieval & Returns cached OSV advisory evidence, affected version ranges, references, and candidate affected features for the finding. \\
\texttt{search\_vulnerable\_feature\_usage} & Vulnerable-feature usage search & Searches repository text for an advisory-derived candidate vulnerable feature; a match is evidence, not proof, of exploitability. \\
\midrule
\multicolumn{3}{l}{\textit{Paper-Context Grounding}} \\
\midrule
\texttt{inspect\_paper\_context} & Paper-context grounding & Returns bounded paper-derived statements about the artifact's intent, threat model, usage, and limitations. \\
\bottomrule
\end{tabularx}
\end{table*}

\section{Tool Usage and Agentic Behavior Analysis}
\label{appxsec:tool_usage}
This subsection analyzes the behavior of the agentic pipeline during repository exploration. We examine the number of reasoning steps and tool calls made by each model, the tools selected from the 20 available options, and the reliability with which runs are completed. This analysis complements the accuracy and cost results above by providing insight into how different models use agentic exploration and how these behaviors relate to their classification performance.

\subsection{Agent Steps and Tool-Call Volume}

\begin{table*}[htbp]
    \centering
    \begin{tabular}{lllccc ccc}
        \toprule
        & & & \multicolumn{3}{c}{\textbf{Agent Steps}} & \multicolumn{3}{c}{\textbf{Tool Calls}} \\
        \cmidrule(lr){4-6} \cmidrule(lr){7-9}
        \textbf{Mode} & \textbf{Model} & & \textbf{Mean} & \textbf{Median (Max)} & \textbf{P90} & \textbf{Mean} & \textbf{Median (Max)} & \textbf{P90} \\
        \midrule
        \multirow{3}{*}{Zero-shot}
        & GPT-5.6 & & 1.00 & 1 (1) & 1.0 & 0.00 & 0 (0) & 0.0 \\
        & GLM-5.2 & & 1.00 & 1 (1) & 1.0 & 0.00 & 0 (0) & 0.0 \\
        & Claude Sonnet 5 & & 0.99 & 1 (1) & 1.0 & 0.00 & 0 (0) & 0.0 \\
        \midrule
        \multirow{3}{*}{Agentic}
        & GPT-5.6 & & 2.35 & 2 (7) & 4.0 & 5.61 & 5 (23) & 12.0 \\
        & GLM-5.2 & & 1.01 & 1 (2) & 1.0 & 0.75 & 0 (184) & 0.0 \\
        & Claude Sonnet 5 & & 2.49 & 2 (12) & 5.0 & 2.58 & 2 (17) & 6.0 \\
        \bottomrule
    \end{tabular}
    \caption{Distribution of agent steps and tool calls per finding across the 250-finding evaluation set. Zero-shot mode performs a single classification pass (agent steps $\approx$1, no tool calls by construction); agentic mode allows up to 12 agent steps. Median (Max) reports the per-finding median with the observed maximum in parentheses; P90 is the 90th percentile.}
    \label{tab:safe_agent_steps}
\end{table*}

\begin{figure}[htbp]
    \centering
    \includegraphics[width=\linewidth]{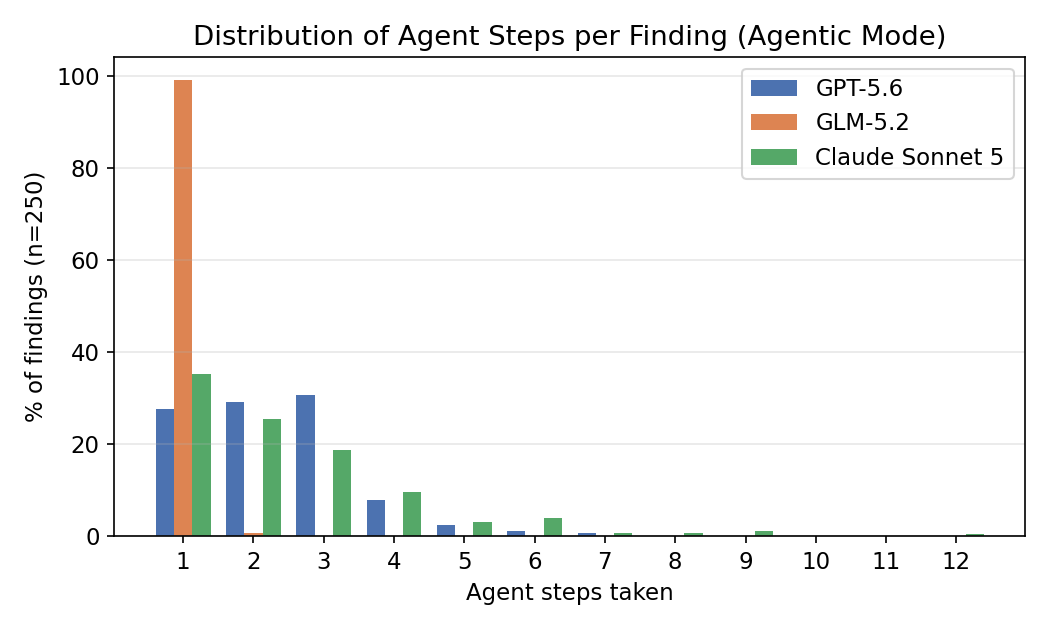}
    \caption{Distribution of agent steps taken per finding in agentic mode, by model.}
    \label{fig:agent_steps_hist}
\end{figure}

Table~\ref{tab:safe_agent_steps} and Figure~\ref{fig:agent_steps_hist} summarize how many agent steps and tool calls each model uses per finding. In zero-shot mode all three models trivially take a single step with zero tool calls by construction. In agentic mode, GPT-5.6 and Claude Sonnet~5 make comparable use of the step budget (mean 2.35 and 2.49 steps respectively, both with a 90th-percentile of 4--5 steps out of the 12 allowed), while GLM-5.2's step count is barely distinguishable from zero-shot mode (mean 1.01 steps, 99\% of findings using exactly one step). The same pattern is sharper in tool-call volume: GPT-5.6 averages 5.61 tool calls per finding (median 5, up to 23), Claude Sonnet~5 averages 2.58 (median 2, up to 17), while GLM-5.2's median is 0 and its mean of 0.75 is driven almost entirely by a single extreme outlier, discussed below.

\subsection{Tool Invocation Patterns}
\begin{table}[htbp]
    \centering
    \begin{tabular}{lcc}
        \toprule
        \textbf{Model (agentic)} & \textbf{$\leq$2 calls} & \textbf{$\geq$3 calls} \\
        \midrule
        GPT-5.6         & 32.4\% & 67.6\% \\
        GLM-5.2         & 99.2\% & 0.8\% \\
        Claude Sonnet 5 & 50.8\% & 49.2\% \\
        \bottomrule
    \end{tabular}
    \caption{Percentage of the 250 agentic findings grouped by tool-call volume.}
    \label{tab:safe_tool_buckets}
\end{table}

\begin{figure}[htbp]
    \centering
    \includegraphics[width=\linewidth]{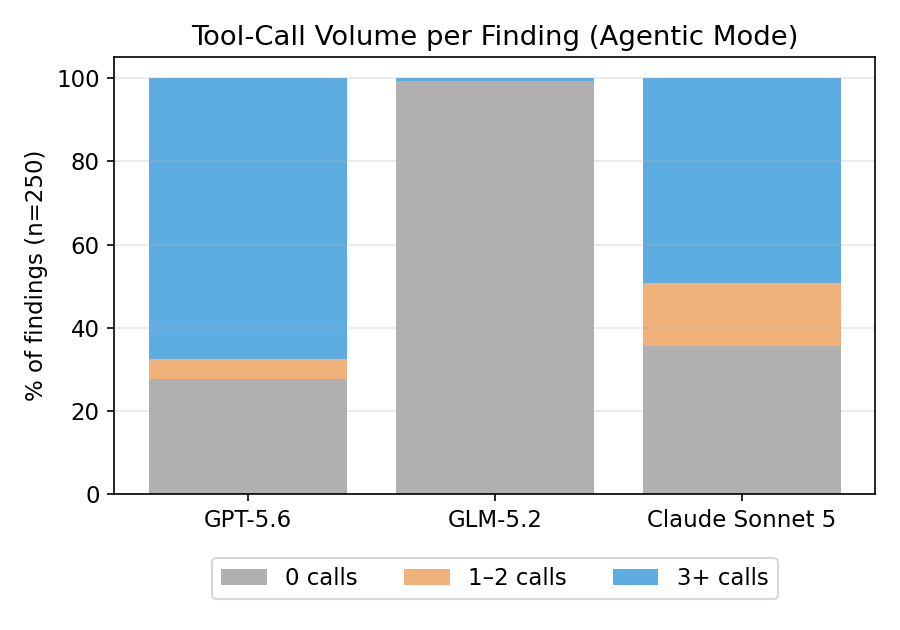}
    \caption{Percentage of agentic findings falling into each tool-call volume bucket, by model.}
    \label{fig:tool_buckets}
\end{figure}

Table~\ref{tab:safe_tool_buckets} and Figure~\ref{fig:tool_buckets} bucket findings by how many tool calls they used. GPT-5.6 makes heavy use of its exploration budget, with 67.6\% of findings using three or more tool calls and only 27.6\% classified with no tool use at all. Claude Sonnet~5 is more selective -- 49.2\% of findings use three or more calls, but a further 35.6\% are classified directly with none -- consistent with a model that decides evidence sufficiency on a per-finding basis rather than always exploring. GLM-5.2 is the outlier: 99.2\% of findings receive zero tool calls, and only 2 of 250 findings (0.8\%) involve any tool use whatsoever, reproducing the figure already noted in the cost analysis above but now isolating it as a distinct behavioral measurement rather than a cost side-effect.

\subsection{Per-Tool Usage Distribution}

\begin{table*}[htbp]
    \centering
    \begin{tabular}{l ccc ccc ccc}
        \toprule
        & \multicolumn{3}{c}{\textbf{GPT-5.6}} & \multicolumn{3}{c}{\textbf{GLM-5.2}} & \multicolumn{3}{c}{\textbf{Claude Sonnet 5}} \\
        \cmidrule(lr){2-4} \cmidrule(lr){5-7} \cmidrule(lr){8-10}
        \textbf{Tool} & \textbf{Calls} & \textbf{\%} & \textbf{Findings} & \textbf{Calls} & \textbf{\%} & \textbf{Findings} & \textbf{Calls} & \textbf{\%} & \textbf{Findings} \\
        \midrule
        search\_repository & 665 & 47.4\% & 149 & 20 & 10.6\% & 2 & 364 & 55.2\% & 139 \\
        read\_file & 347 & 24.8\% & 112 & 89 & 47.3\% & 1 & 78 & 11.8\% & 45 \\
        read\_snippet & 58 & 4.1\% & 30 & 57 & 30.3\% & 1 & 11 & 1.7\% & 9 \\
        search\_vulnerable\_feature\_usage & 78 & 5.6\% & 36 & 3 & 1.6\% & 1 & 28 & 4.2\% & 24 \\
        inspect\_vulnerability\_advisory & 43 & 3.1\% & 43 & 1 & 0.5\% & 1 & 29 & 4.4\% & 29 \\
        inspect\_package\_usage & 59 & 4.2\% & 57 & 1 & 0.5\% & 1 & 5 & 0.8\% & 5 \\
        inspect\_entrypoints & 29 & 2.1\% & 29 & 1 & 0.5\% & 1 & 17 & 2.6\% & 17 \\
        list\_repository\_tree & 18 & 1.3\% & 18 & -- & -- & -- & 24 & 3.6\% & 24 \\
        find\_python\_callers & 30 & 2.1\% & 15 & 2 & 1.1\% & 1 & 8 & 1.2\% & 5 \\
        inspect\_python\_imports & 6 & 0.4\% & 4 & 11 & 5.9\% & 1 & 23 & 3.5\% & 11 \\
        inspect\_dependency\_files & 22 & 1.6\% & 22 & -- & -- & -- & 15 & 2.3\% & 15 \\
        find\_symbol\_references & 16 & 1.1\% & 12 & -- & -- & -- & 17 & 2.6\% & 14 \\
        inspect\_paper\_context & 4 & 0.3\% & 4 & -- & -- & -- & 21 & 3.2\% & 21 \\
        inspect\_dependency\_graph & 14 & 1.0\% & 14 & 1 & 0.5\% & 1 & 2 & 0.3\% & 2 \\
        find\_python\_definitions & 3 & 0.2\% & 3 & 1 & 0.5\% & 1 & 11 & 1.7\% & 10 \\
        inspect\_python\_function & 2 & 0.1\% & 1 & -- & -- & -- & 6 & 0.9\% & 4 \\
        find\_python\_security\_operations & 6 & 0.4\% & 6 & 1 & 0.5\% & 1 & 1 & 0.2\% & 1 \\
        inspect\_python\_scope & 1 & 0.1\% & 1 & -- & -- & -- & -- & -- & -- \\
        trace\_python\_variable & 1 & 0.1\% & 1 & -- & -- & -- & -- & -- & -- \\
        trace\_python\_reachability & -- & -- & -- & -- & -- & -- & -- & -- & -- \\
        \midrule
        \textbf{Total tool calls} & \textbf{1402} & -- & -- & \textbf{188} & -- & -- & \textbf{660} & -- & -- \\
        \bottomrule
    \end{tabular}
    \caption{Usage of each of the 20 tools available to the agentic pipeline, counted only from each finding's final (successful) classification attempt. Calls is the raw invocation count across the 250-finding set; \% is that tool's share of the model's total tool calls; Findings is the number of distinct findings (of 250) in which the tool was invoked at least once. 
    }
    \label{tab:safe_tool_distribution}
\end{table*}

\begin{figure}[htbp]
    \centering
    \includegraphics[width=\linewidth]{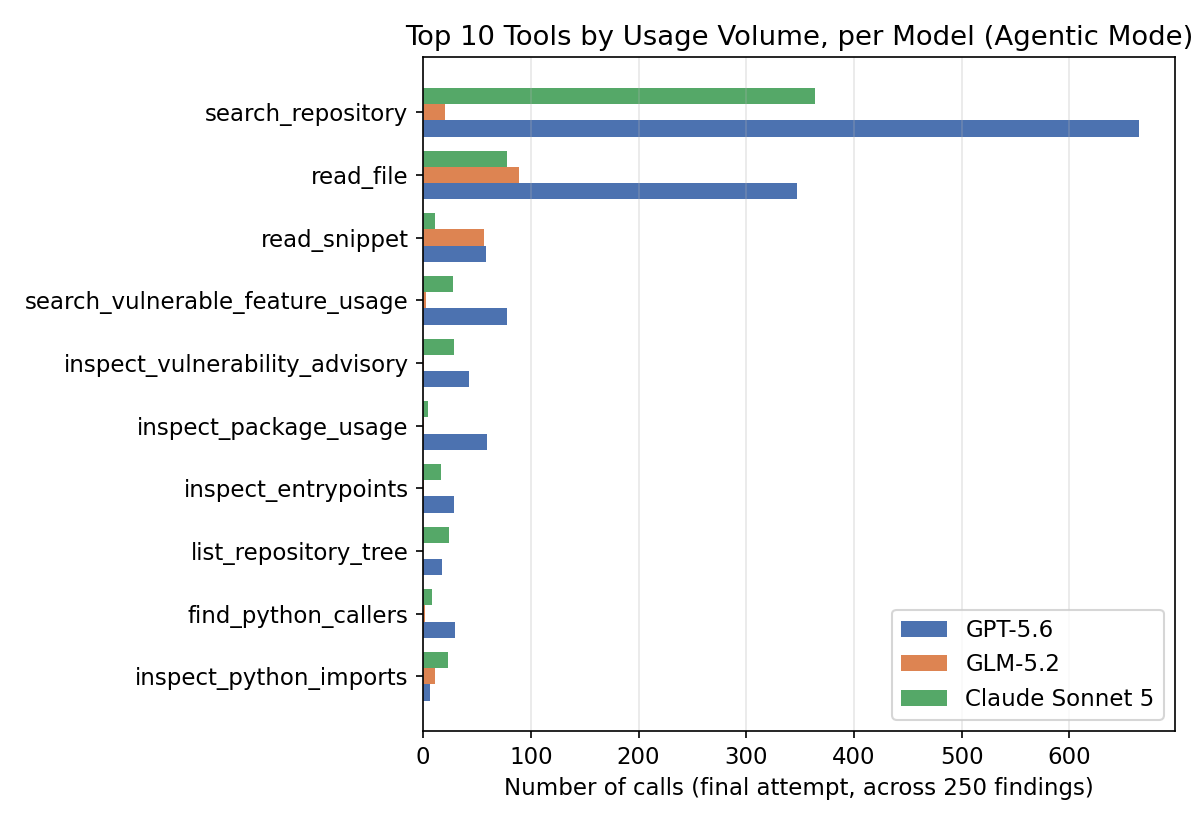}
    \caption{The 10 most-used tools (by total call volume, summed across models) and each model's usage of them.}
    \label{fig:top_tools}
\end{figure}

Table~\ref{tab:safe_tool_distribution} and Figure~\ref{fig:top_tools} break tool use down by the specific tool invoked, counted from each finding's final (successful) classification attempt. For both GPT-5.6 and Claude Sonnet~5, \texttt{search\_repository} is the dominant tool (47.4\% and 55.2\% of all their tool calls, respectively), used in the majority of their tool-using findings (149 and 139 findings) as a first-pass discovery step, typically followed by \texttt{read\_file} to inspect a located call site. Beyond these two, the two models diverge somewhat in emphasis: GPT-5.6 makes comparatively more use of \texttt{inspect\_package\_usage} (59 calls, 57 distinct findings) and \texttt{search\_vulnerable\_feature\_usage} (78 calls, 36 findings), while Claude Sonnet~5 makes distinctively more use of \texttt{inspect\_paper\_context} (21 calls, 21 findings) and \texttt{list\_repository\_tree} (24 calls, 24 findings), suggesting a relatively greater emphasis on situating a finding within the paper's stated research intent and the repository's overall layout before drilling into a specific file. Of the 20 available tools, \texttt{trace\_python\_reachability} is never invoked by any model across all 750 agentic findings, and several of the fine-grained static-analysis tools (\texttt{inspect\_python\_scope}, \texttt{trace\_python\_variable}, \texttt{inspect\_python\_function}) are used only a handful of times in total, indicating that in practice models rely on a small core of coarse-grained discovery and read tools rather than the more specialized reachability- and scope-tracing tools built into the pipeline.

GLM-5.2's tool usage is not only rare but concentrated: of its 188 total tool calls, 184 come from a single finding (\texttt{SP2023-artifacts/A013}, a \texttt{trivy-vulnerability} finding for CVE-2025-2148), where the model issued a dense burst of \texttt{read\_file} (89 calls) and \texttt{read\_snippet} (57 calls) invocations within just two agent steps rather than across many; the second and only other tool-using finding contributes the remaining 4 calls. Every other tool in Table~\ref{tab:safe_tool_distribution} shows GLM-5.2 touching it in exactly one finding. In other words, GLM-5.2's apparent per-tool ``coverage'' of the catalog is an artifact of one atypical finding exercising most of the tool surface at once, not evidence of broad, routine tool engagement -- reinforcing that GLM-5.2's default behavior, in 248 of 250 findings, is to classify directly from the base evidence packet exactly as it would in zero-shot mode.

\subsection{Run Reliability}

\begin{table*}[htbp]
    \centering
    \begin{tabular}{lllccc}
        \toprule
        \textbf{Mode} & \textbf{Model} & & \textbf{Findings} & \textbf{Succeeded} & \textbf{Never Succeeded} \\
        \midrule
        \multirow{3}{*}{Zero-shot}
        & GPT-5.6 & & 250 & 250 & 0 \\
        & GLM-5.2 & & 250 & 250 & 0 \\
        & Claude Sonnet 5 & & 250 & 248 & 2 \\
        \midrule
        \multirow{3}{*}{Agentic}
        & GPT-5.6 & & 250 & 250 & 0 \\
        & GLM-5.2 & & 250 & 250 & 0 \\
        & Claude Sonnet 5 & & 250 & 249 & 1 \\
        \bottomrule
    \end{tabular}
    \caption{Run-level reliability across the 250-finding evaluation set. Never Succeeded denotes findings for which no successful classification was obtained and which are therefore excluded from that run's evaluated predictions.}
    \label{tab:safe_reliability}
\end{table*}

Table~\ref{tab:safe_reliability} summarizes run-level reliability across models and evaluation modes. GPT-5.6 and GLM-5.2 successfully classify all 250 findings in both zero-shot and agentic modes. Claude Sonnet~5 fails to produce a successful classification for 2 findings in zero-shot mode and 1 finding in agentic mode due to schema-parse failures. These findings are excluded from the corresponding evaluated predictions, consistent with the ``no-prediction'' counts reported in the classification results above.

\subsection{Runtime}

\begin{figure}[htbp]
    \centering
    \includegraphics[width=\linewidth]{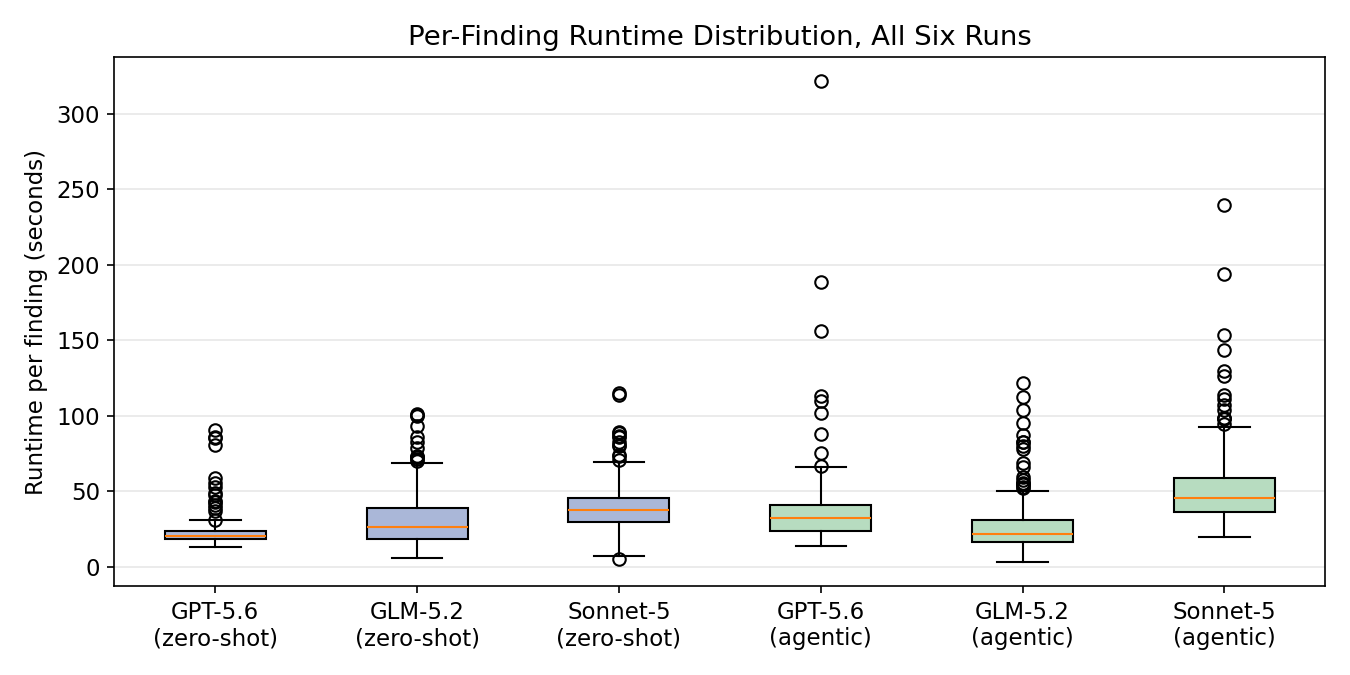}
    \caption{Per-finding runtime distribution across all runs.
    }
    \label{fig:runtime_boxplot}
\end{figure}

Figure~\ref{fig:runtime_boxplot} shows the per-finding runtime distribution across all six runs. Runtime tracks tool-call volume closely: GPT-5.6's agentic median runtime (32.2s) is 1.6$\times$ its zero-shot median (20.6s), and Claude Sonnet~5's agentic median (45.5s) is 1.2$\times$ its zero-shot median (37.5s), both consistent with the additional model round-trips per finding shown in Table~\ref{tab:safe_agent_steps}. GLM-5.2's agentic median runtime (21.6s) is, by contrast, \emph{lower} than its own zero-shot median (26.1s), a further indication that its agentic run is not, in practice, doing more work than its zero-shot pass. All three models show a long right tail of slow outliers in agentic mode (up to 322s for GPT-5.6, 239s for Claude Sonnet~5), corresponding to the minority of findings that exhaust most of the available agent-step budget.

\section{Permanent Classification Failures}
\label{app:permanent_failures}

Table~\ref{tab:safe_permanent_failures} lists all permanently failed findings observed across the evaluation runs. 
There are exactly three, and all three occur on Claude Sonnet~5, two in zero-shot mode and one in agentic mode, while GPT-5.6 and GLM-5.2 have zero permanent failures in either mode across runs. A striking pattern in these three failures is that all of them are \texttt{trivy} findings. The two zero-shot failures share an identical root cause that the model returned an empty, non-JSON response on all three schema-repair attempts, while the agentic failure has a different cause: the model spent its entire 12-step budget exploring the repository and data without ever emitting a final classification. With only three occurrences this pattern is suggestive rather than conclusive, but it points to \texttt{trivy} findings' longer or more complex evidence packets as a plausible stressor on Claude Sonnet~5's response formatting and step-budget management specifically, since the same finding type causes no failures at all for GPT-5.6 or GLM-5.2.
\begin{table*}[htbp]
    \centering
    \caption{Permanently-failed finding (no successful classification in any attempt) across all runs. 
    }
    \label{tab:safe_permanent_failures}
    \begin{tabular}{lll p{5.2cm} l}
        \toprule
        \textbf{Mode} & \textbf{Model} & \textbf{Finding (artifact / tool / CVE-ID)} & \textbf{Failure mode} & \textbf{Attempts} \\
        \midrule
        Zero-shot & Claude Sonnet 5 & A335 / trivy / CVE-2025-2149 & Empty/unparseable model response (JSON schema attempts exhausted) & 3/3 schema attempts \\
        Zero-shot & Claude Sonnet 5 & A081 / trivy / CVE-2024-35195 & Empty/unparseable model response (JSON schema attempts exhausted) & 3/3 schema attempts \\
        Agentic & Claude Sonnet 5 & A044 / trivy / CVE-2026-59204 & Exhausted 12-step agent budget without a final classification & 12/12 agent steps \\
        \bottomrule
    \end{tabular}
\end{table*}

\section{Case Study 1 Output}

\begin{lstlisting}[language=json,caption={Case Study 1: SAFE output, agentic mode (A078)},label={lst:casestudy1_output}]
{
  (*@{\color{blue}"security\_label":}@*) "CONTEXTUAL_RISK",
  (*@{\color{blue}"code\_purpose":}@*) "Uploads the generated Liquibase convenience ZIP as a workflow artifact....supports the CI/package process for a cache-leakage proof-of-concept....mutable action resolution is not itself the vulnerability being studied.",
  (*@{\color{blue}"execution\_context":}@*) "Executes third-party Action code on a hosted CI runner during the package job....signing-key material is decoded into workspace files under liquibase-dist/target/keys before the flagged action runs from the same workspace.",
  (*@{\color{blue}"required\_conditions\_for\_exploit":}@*) "....workflow must reach the package job....GitHub must resolve the mutable v3 reference....attacker must have compromised the action publisher or gained ability to repoint that tag before resolution....",
  (*@{\color{blue}"input\_controlled\_by\_attacker":}@*) "yes - an attacker who compromises the action publisher or repoints the upstream v3 tag controls the fetched action code....ordinary contributors do not control this reference.",
  (*@{\color{blue}"reachable\_in\_artifact\_execution":}@*) "yes -....active step in the package workflow....runs via workflow_dispatch, pull_request_target, and pushes to master/main/github-action-* branches....README documents triggering on a github-action-* branch.",
  (*@{\color{blue}"evidence\_snippet":}@*) "....build.yml:375-392 decodes signing-key material into workspace files....build.yml:460 contains uses: actions/upload-artifact@v3, unpinned....",
  (*@{\color{blue}"reasoning":}@*) "....scanner-matched condition is present and reachable....mutable action tag is not the cache-leakage mechanism being studied; it is an incidental supply-chain weakness....action is resolved across a live external trust boundary, and its code runs after sensitive files have been created in the workspace....contextual risk rather than benign research usage or a mere hardening observation.",
  (*@{\color{blue}"recommendation":}@*) "....pin every action reference to a reviewed commit SHA....apply least-privilege workflow permissions....avoid leaving signing keys in the workspace across third-party action steps...."
}
\end{lstlisting}

\section{Case Study 2 Output}

\begin{lstlisting}[language=json,caption={Case Study 2: SAFE output, Claude Sonnet 5 agentic (A088)},label={lst:casestudy2_output}]
{
  (*@{\color{blue}"security\_label":}@*) "CONTEXTUAL_RISK",
  (*@{\color{blue}"code\_purpose":}@*) "requirements.txt pins application/runtime dependencies for an LLaVA-based email-assistant simulation studying the RAGworm self-replicating-prompt attack....flagged urllib3 pin is a supporting HTTP-stack dependency used to fetch remote images referenced in processed content.",
  (*@{\color{blue}"execution\_context":}@*) "Locally-deployed research prototype....LLaVaServer component receives images potentially referenced by attacker-influenced URLs from RAG/email content and fetches them via requests/urllib3....explicitly modeled to receive content from other (attacker-simulated) parties.",
  (*@{\color{blue}"required\_conditions\_for\_exploit":}@*) "....pinned vulnerable urllib3 (2.2.1, <2.6.3) remains installed....attacker supplies or controls a URL that load_image() fetches....target server issues a redirect with a maliciously compressed body and redirects are followed (default behavior)....",
  (*@{\color{blue}"input\_controlled\_by\_attacker":}@*) "yes - Run_LLaVa.py:75-81 defines load_image(image_file), invoked at line 171 within a documented entrypoint....per the artifact's own threat model (RAGworm), image references processed by this path originate from email/RAG content treated as attacker-controllable.",
  (*@{\color{blue}"reachable\_in\_artifact\_execution":}@*) "yes - requests (pinned in the same manifest) relies on urllib3 and streams responses while following redirects by default....load_image() is reachable from the documented LLaVaServer entrypoint.",
  (*@{\color{blue}"evidence\_snippet":}@*) "....requirements.txt line 305 (UTF-16, confirmed via raw byte-level read): urllib3==2.2.1....Run_LLaVa.py lines 75-81: load_image() calls requests.get(image_file) for any http/https-prefixed reference, called at line 171....",
  (*@{\color{blue}"reasoning":}@*) "....UTF-16 encoding defeated the automated declared/imported check, a tooling gap rather than proof of absence....artifact contains a concrete, documented code path that performs unrestricted requests.get() calls on URLs plausibly attacker-influenced per the paper's own stated threat model....real, version-confirmed vulnerability + documented plausible attacker-influenced path + incidental research intent supports CONTEXTUAL_RISK.",
  (*@{\color{blue}"recommendation":}@*) "....upgrade urllib3 to >=2.6.3....disable automatic redirect following or set explicit response-size limits/timeouts when fetching remote content from any source that may be attacker-influenced....re-verify installed dependency versions, since the manifest's UTF-16 encoding should also be fixed...."
}
\end{lstlisting}

\end{document}